\documentclass[twocolumn,tight,times]{aastex631}

\usepackage{graphicx}	
\usepackage{color,soul}

\usepackage{amsmath}	
\usepackage{amssymb}	
\usepackage{amsbsy}	
\usepackage{newtxtt,newtxmath}
\usepackage[T1]{fontenc}
\usepackage{catchfile}
\usepackage{enumitem}

	\newcommand{\lya}{Ly$\alpha$}	
	\newcommand{\halpha}{H$\alpha$}
	\newcommand{\hbeta}{H$\beta$}
	\newcommand{\hgamma}{H$\gamma$}
	
	\newcommand{\heii}{He{\sc ii}}
	\newcommand{\nii}{[N{\sc ii}]}
	\newcommand{\oii}{[O{\sc ii}]}
	\newcommand{\oiii}{[O{\sc iii}]}
	\newcommand{\neiii}{[Ne{\sc iii}]}
	
	\newcommand{\cgsline}{erg s$^{-1}$ cm$^{-2}$}
	
	\newcommand{\msol}{M$_\odot$}
	\newcommand{\zsol}{$Z_\odot$}	
	\newcommand{\pypeit}{\texttt{Pypeit}}
	\newcommand{\pyqsofit}{\texttt{PyQSOFit}}
	\newcommand{\pyneb}{\texttt{PyNeb}}
	\newcommand{\xiion}{$\xi_{ion}$}
	\newcommand{\xiionHtwo}{$\xi_{ion}^\textrm{H{\sc ii}}$}
	
	\newcommand{\cigale}{\texttt{Cigale}}
	\newcommand{\bagpipes}{\texttt{Bagpipes}}
	\newcommand\C[1]\null

\begin{document}
			
			\title{EELG1002: A Record-Breaking \oiii+\hbeta~EW $\sim3700$\AA~Galaxy at $z \sim 0.8$ -- Analog of Early Galaxies?}

			\correspondingauthor{Ali Ahmad Khostovan}
			\email{akhostov@gmail.com}
			
			\author[0000-0002-0101-336X]{Ali Ahmad Khostovan}
			\affiliation{Laboratory for Multiwavelength Astrophysics, School of Physics and Astronomy, Rochester Institute of Technology, 84 Lomb Memorial Drive, Rochester, NY 14623, USA}
			\affiliation{Department of Physics and Astronomy, University of Kentucky, 505 Rose Street, Lexington, KY 40506, USA}

			\author[0000-0001-9187-3605]{Jeyhan S. Kartaltepe}
			\affiliation{Laboratory for Multiwavelength Astrophysics, School of Physics and Astronomy, Rochester Institute of Technology, 84 Lomb Memorial Drive, Rochester, NY 14623, USA}

			\author[0000-0002-0245-6365]{Malte Brinch}
			\affiliation{Cosmic Dawn Center (DAWN), Denmark}
			\affiliation{DTU-Space, National Space Institute, Technical University of Denmark, Elektrovej 327, DK-2800 Kgs. Lyngby, Denmark}
	
			\author[0000-0002-0930-6466]{Caitlin Casey}
			\affiliation{The University of Texas at Austin, 2515 Speedway Blvd Stop C1400, Austin, TX 78712, USA}	
			\affiliation{Cosmic Dawn Center (DAWN), Denmark}

			\author[0000-0002-9382-9832]{Andreas Faisst}
			\affiliation{Caltech/IPAC, 1200 E. California Blvd., Pasadena, CA, 91125 USA}
			
			\author[0000-0003-0129-2079]{Santosh Harish}
			\affiliation{Laboratory for Multiwavelength Astrophysics, School of Physics and Astronomy, Rochester Institute of Technology, 84 Lomb Memorial Drive, Rochester, NY 14623, USA}

			\author[0000-0002-0236-919X]{Ghassem Gozaliasl}
			\affiliation{Department of Computer Science, Aalto University, PO Box 15400 Espoo, 00 076 Finland}
			\affiliation{Department of Physics, University of Helsinki, PO Box 64 00014 Helsinki, Finland}

			\author[0000-0003-3228-7264]{Masato Onodera}
			\affiliation{Subaru Telescope, National Astronomical Observatory of Japan, National Institutes of Natural Sciences (NINS), 650 North A'ohoku Place, Hilo, HI 96720, USA}
			\affiliation{Department of Astronomical Science, The Graduate University for Advanced Studies, SOKENDAI, 2-21-1 Osawa, Mitaka, Tokyo 181-8588, Japan}
	
			\author[0000-0001-6229-4858]{Kiyoto Yabe}
			\affiliation{Subaru Telescope, National Astronomical Observatory of Japan, National Institutes of Natural Sciences (NINS), 650 North A'ohoku Place, Hilo, HI 96720, USA}

			

		\begin{abstract}
		Extreme emission line galaxies (EELGs) are powerful low-$z$ analogs of high-$z$ galaxies that can provide us valuable insights of early Universe conditions. We present a detailed analysis of EELG1002: a $z=0.8275$ EELG identified within archival Gemini/GMOS spectroscopy as part of the on-going COSMOS Spectroscopic Archive. We find EELG1002 is a low-mass ($\sim10^{8}$ \msol), compact ($\sim530$ pc), bursty star-forming galaxy with a $\sim15-35$ Myr mass doubling timescale. EELG1002 has record-breaking rest-frame \oiii+\hbeta~EW $\sim3100-3700$\AA; $\sim32-36\times$ higher than typical $z \sim 0.8$ \oiii~emitters with similar stellar mass and higher than typical $z > 5$ galaxies. We find no clear evidence of an AGN suggesting the emission lines are star formation driven. EELG1002 is chemically unevolved (direct $T_e$; $12+\log_{10}(\textrm{O/H})\sim7.52$ consistent with $z>5$ galaxies at fixed stellar mass) and may be undergoing a first intense, bursty star formation phase analogous to conditions expected of galaxies in the early Universe. We find evidence for a highly energetic ISM (\oiii/\oii~$\sim9$) and hard ionizing radiation field (elevated \neiii/\oii~at fixed \oiii/\oii). Coupled with its compact, metal-poor, and actively star-forming nature, EELG1002 is found to efficiently produce ionizing photons ($\textrm{\xiion}\sim10^{25.74}~$erg$^{-1}$ Hz) and may have $\sim10-20\%~$LyC escape suggesting such sources may be important analogs of galaxies responsible for reionization. We find dynamical mass of $\sim10^9~$\msol~suggesting copious amounts of gas to support intense star formation as also suggested by identified Illustris-TNG analogs. EELG1002 may be an ideal low-$z$ laboratory of galaxies in the early Universe and demonstrates how archival datasets can support high-$z$ science and next-generation surveys planned with \textit{Euclid} and \textit{Roman}.
		\end{abstract}
		
            \keywords{\href{http://astrothesaurus.org/uat/594}{Galaxy Evolution (594)}, \href{http://astrothesaurus.org/uat/734}{High-redshift Galaxies (734)}, \href{http://astrothesaurus.org/uat/847}{Interstellar medium (847)}, \href{http://astrothesaurus.org/uat/1570}{Starburst Galaxies (1570)}, \href{http://astrothesaurus.org/uat/1569}{Star Formation (1569)}}

		
		
		\section{Introduction}
            Understanding how star-formation occurred in the early Universe is crucial in our understanding of how galaxy formation and cosmic Reionization occurred. Prior to \textit{JWST}, it was very difficult to observe galaxies at $z > 6$ and was primarily limited to samples selected as Lyman Break Galaxies (LBGs; e.g., \citealt{Bouwens2011,McLure2013,Oesch2014,McLeod2016,Ishigaki2018,Finkelstein2022}), \lya~emitters (e.g., \citealt{Rhoads2000,Malhotra2004,Dawson2007,Matthee2014,Santos2016,Konno2018,Sobral2018,Taylor2020,Goto2021,Wold2022,Torralba2024}), and galaxies selected based on nebular excess within the \textit{Spitzer}/IRAC bands (e.g., \citealt{Shim2011,Smit2015,Faisst2016_EW,Marmol2016,Rasappu2016,deBarros2019,Lam2019,Endsley2021_EW}). However, in only the first few years of \textit{JWST} we are detecting numerous $z > 6$ galaxies that were missed in past selection techniques and also galaxies with redshifts in the double digits. \textit{JWST} has pushed our observable window further towards the era that we expect harbors the first generation of galaxies. However, local analogs of high-$z$ galaxies can still provide us with valuable insight on the conditions of galaxies in the early Universe that are still limiting with even \textit{JWST}. 
            
            Extreme Emission Line Galaxies (EELGs) are a unique subset of galaxies known for having strong nebular emission line features with high equivalent widths (EWs; ratio of line flux to continuum flux density) at a level that can dominate some of the widest broadband filters. Other names associated with EELGs include `Green Peas' ($z \sim 0.2 - 0.3$; \citealt{Cardamone2009,Izotov2011}) and `Blueberries' ($z < 0.05$; \citealt{Yang2017_BB}) given their strong nebular contribution in the SDSS $r$ and $g$ bands, respectively. EELGs are known to be low-mass systems (e.g., \citealt{Maseda2014,Amorin2015,Forrest2017,Yang2017_BB}) with compact sizes (e.g., \citealt{vanderWel2011,Amorin2015,Forrest2017,Yang2017_GP_sizes,Kim2021}) typical of galaxies at $z > 3$ (e.g., \citealt{Yang2022_Lilan,Ormerod2024}), and with elevated star-formation rates and mass doubling timescales of $< 100$ Myr (e.g., \citealt{Atek2011, Maseda2013, Atek2014, Amorin2015}). Past studies reveal that EELGs are low in gas-phase metallicities (e.g., \citealt{Amorin2014_VUDS,Amorin2014_highz,Jiang2019}) with energetic ISM conditions (elevated \oiii/\oii~ratios; e.g., \citealt{Izotov2018,Paalvast2018,Izotov2021_O32}). EELGs also exhibit LyC escape with high ionizing photon production efficiencies (\xiion) owing to their compact, star-forming nature and high \oiii/\oii~ratios (e.g., \citealt{Schaerer2016,Tang2019,Emami2020,Atek2022}). This makes EELGs powerful analogs of sources that contribute towards cosmic reionization in the high-$z$ Universe where measurements of LyC escape are extremely difficult given the IGM transmission at $z > 3$ (e.g., \citealt{Madau1995,Inoue2014}). It also allows for detailed analysis on the conditions and mechanisms associated with LyC escape that are expected to occur during the Epoch of Reionization.		
            
             Strong nebular features (e.g., \oiii, \hbeta, \halpha) confirm recent and intense star-formation activity. \cite{Tang2022} used non-parametric star formation history modeling and found that EELGs not only go through a recent, intense burst of star-formation but may have also gone through past bursts highlighting episodic SFHs. \cite{Cohn2018} suggests that high EW systems at high-$z$ represent galaxies undergoing a first bursty phase of star-formation activity given that stellar mass buildup has not fully taken effect (fainter continuum flux densities) resulting in higher \oiii+\hbeta~EW. After each subsequent burst, the stellar continuum increases in brightness resulting in lower EWs. Hydrodynamical simulations and simple analytical models also suggest that star formation activity in high-$z$ galaxies is burst-dominated (e.g.; \citealt{Sparre2017,Faucher2018}). Given that high EWs are ubiquotous of high-$z$ galaxy populations (e.g., \citealt{Smit2014,Khostovan2024}), identifying high EW EELGs in the low-$z$ Universe provides an interesting window in studying the star formation processes expected to occur in the high-$z$ Universe. Coupled with past confirmations of LyC escape and elevated \xiion, low-$z$ EELGs are a window to explore how star formation processes and ionizing photon production of low-mass, star-forming galaxies can contribute towards the cosmic ionizing photon budget needed to facilitate cosmic reionization. 
            
            In this paper, we present a detailed analysis of EELG1002 ($\alpha = 10:00:32.304$, $\delta = +2:51:11.351$): a record-breaking $z \sim 0.8$ EELG that was serendipitously identified within archival Gemini/GMOS spectroscopy as part of ongoing work in developing the COSMOS Spectroscopic Archive. This source has properties that are strongly consistent with some of the most extreme star-forming galaxies currently being observed with \textit{JWST} at $z > 6$ and provides an ideal low-$z$ laboratory to investigate the star-formation, ISM, and ionizing properties of galaxies that exist in the high-$z$ Universe. In this work, we investigate EELG1002 in great detail using a combination of GMOS spectra, detailed spectrophotometric SED fitting, morphology measurements, and analogs within hydrodynamical simulations to investigate the nature of this source within the context of star-formation processes, ISM conditions, and potential as an important contributor of ionizing photons. The main objective of this paper is to first highlight the importance of such a high EW EELG and motivate for further search of EELG1002-like systems to develop larger statistical samples for analysis. Second, we aim to demonstrate the power of using archival data in finding such hidden gems that can enable new science and support future science objectives. 
            
            The structure of this paper is as follows: \S\ref{sec:data} outlines the Gemini/GMOS observations of EELG1002 along with the ancillary data from COSMOS2020 \citep{Weaver2022} and additional \textit{HST}/WFC3 F140W imaging \citep{Silverman2018,Ding2020}. \S\ref{sec:methodology} presents our approach in reducing the spectroscopic data, determining the spectroscopic redshift, emission line profile fitting, and measuring the velocity dispersion, ISM properties, EWs, ionizing photon production efficieny, sizes, and SED fitting procedure. We present all our results in \S\ref{sec:results} using all available spectroscopic and photometric evidence to characterize the nature of EELG1002 and how it is analogous to $z > 6$ galaxies currently being observed with \textit{JWST}. We present further discussion in \S\ref{sec:discussion} in regards to the star-formation history showing how it is supported by elevated gas masses by using both measurements of dynamical masses and EELG1002-like sources within Illustris-TNG. We also discuss the descendants of sources like EELG1002, the feasibility of LyC escape, and how EELG1002 may be an ideal case of studying reionization-era galaxies in the low-$z$ Universe. \S\ref{sec:conclusions} outlines the main conclusions of this paper.

            Throughout this paper, we assume a $\Lambda$CDM cosmology with $H_0 = 70$ km s$^{-1}$ Mpc$^{-1}$, $\Omega_m = 0.3$ and $\Omega_\Lambda = 0.7$. All magnitudes, unless otherwise stated, follow the AB magnitude system.
            	
		\section{Data}
		\label{sec:data}
		
		The COSMOS Spectroscopic Archive (Khostovan et al., in prep) is a great effort of gathering and processing all ground-based spectroscopic observations done over the past two decades with the COSMOS legacy field \citep{Scoville2007}. As part of our search, we came across EELG1002 within the Gemini Science Archive which was part of a Fast Turnaround program, GS-2017A-FT-9 (PI: Kiyoto Yabe). The main focus of the program was to target emission line galaxies at $z > 0.3$ that were identified by strong nebular excess within the Subaru/HSC broadband filters indicative of high EW emission lines. These sources are expected to have strong \oiii~EW of $>2000$\AA~and are metal-poor with the possibility of \oiii4363\AA~detection. 
		
		\subsection{GMOS Spectroscopy}
		
		\begin{figure}
			\centering
			\includegraphics[width=\columnwidth]{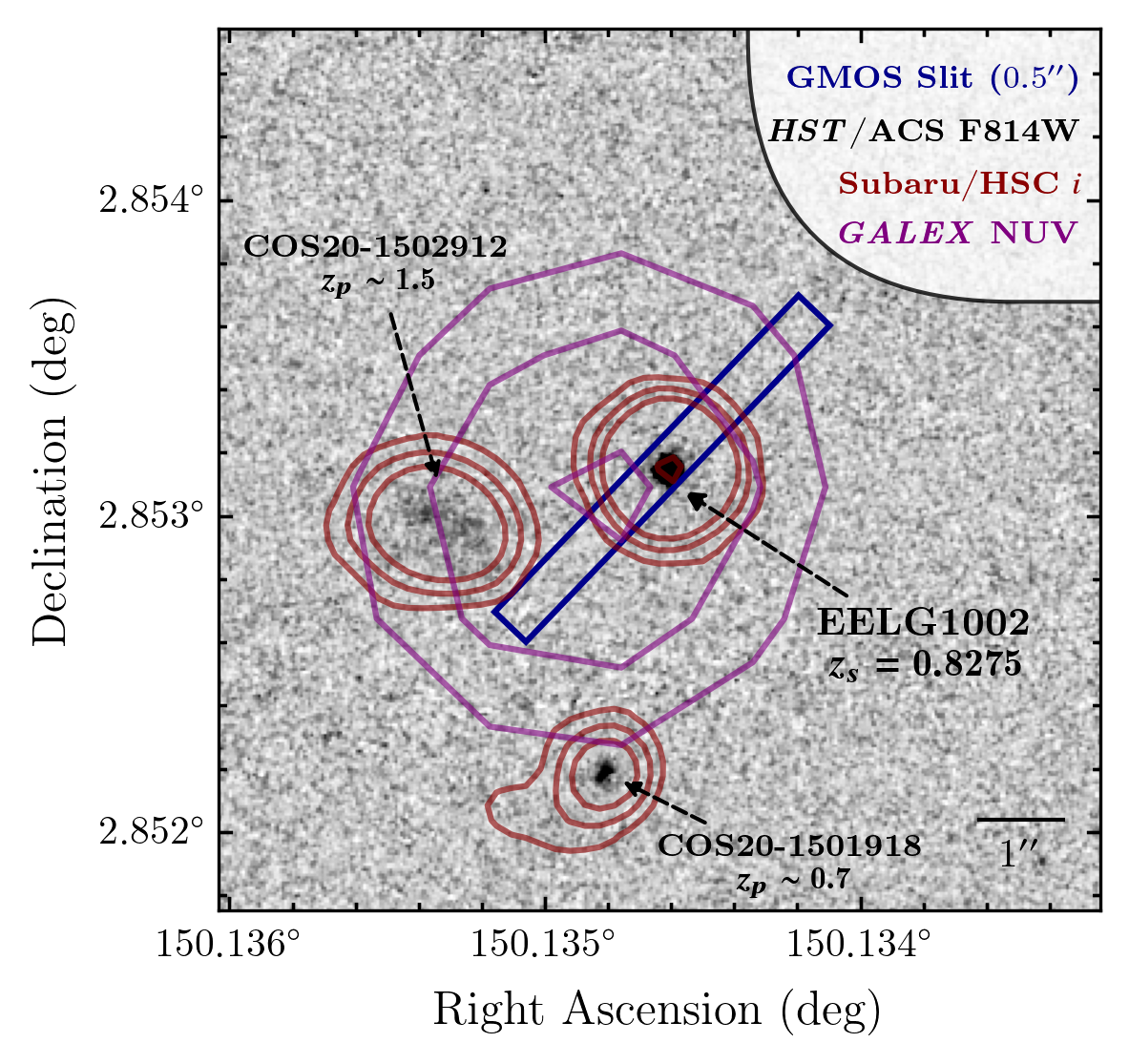}
			\caption{GMOS slit position (\textit{blue}) for EELG1002 with a $10'' \times 10''$ \textit{HST}/ACS F814W cutout ($0.03''$ pix$^{-1}$ resolution; \textit{background}). Subaru/HSC $i$ detection is shown in \textit{red} contours highlighting the lower resolution of ground-based observations where we find $\sim 50$\% of EELG1002's total flux is observed along the $0.5''$ GMOS slit. EELG1002 has \textit{GALEX}/NUV (\textit{purple} contours) and FUV detections which would suggest strong \lya~emission and possible LyC escape, respectively; however, \textit{GALEX} spatial resolution is quite poor. The proximity of two nearby sources along line-of-sight, shown with their COSMOS2020/Classic IDs and best-fit photometric redshifts, suggests the \textit{GALEX} detections suffer blending issues. We therefore ignore \textit{GALEX}, as well as \textit{Spitzer}, photometry in this analysis due to blending issues.}
			\label{fig:slitpos}
		\end{figure}
	
		GS-2017A-FT-9 consisted of three masks with only one in COSMOS which is the focus of this paper. Observations were taken on 23 April 2017 with Gemini-South using the Gemini Multiobject Spectrograph (GMOS) and its Hamamatsu Detector, R150 grating, and GG455 blocking filter. This allows for spectral coverage from 4600\AA~to 10000\AA. Seeing conditions were reported as $0.9''$ throughout the observing time and humidity levels were somewhat elevated at $\sim 40\%$. A total of 3 individual exposures each having a different dispersion angle were pointed at GS2017AFT009-02 (mask design file) with the central wavelengths of each exposure being 8070\AA, 8200\AA, and 8330\AA, respectively, which when 2D coadded would take into account the GMOS detector chip gaps allowing for continuous spectral coverage in the final 1D and 2D spectra. Each science exposure was 1200 seconds resulting in an on-source integration time of 3600 seconds. The slit widths were set to $0.5''$ for the science targets and box slits of $2''$ sizes were used for alignment stars. A single 5-sec twilight flat and 1-sec dome flat was taken per dispersion angle totaling 3 twilight flats and 3 dome flats. A single 20-sec CuAr arc lamp exposure was also taken for wavelength calibrations per each dispersion angle. The bias frames were taken on 29 April 2017 (the last night of the program) and consisted of 5 frames each 1 second in exposure time.

		The standard star LTT7379 (G0 star) that we used for flux calibration purposes was observed on 29 April 2017 with the same configuration and central wavelength of 8200\AA. Seeing was reported as $1.08''$ with humidity levels of $14\%$. Observations of the standard star were performed using a $0.5''$ wide long-slit for an on-source integration time of 10 seconds. A single 1-sec CuAr arc lamp and dome flat were observed with the same configuration and slit widths.
		
		\subsection{COSMOS2020 Classic}
		We cross-match EELG1002 with the COSMOS2020/Classic catalog \citep{Weaver2022} which has high-quality, multi-wavelength photometry enabling us to investigate its spectral energy distribution (SED). A total of 38 photometric bands are included from space-based missions (\textit{GALEX}, \textit{HST}, \textit{Spitzer}) and ground-based observatories (CFHT/MegaCam, Subaru/SuprimeCam, Subaru/HyperSuprimeCam, Paranal/VISTA). The available photometry used in this work is highlighted in Table \ref{table:photometry}. We choose to ignore \textit{GALEX} and \textit{Spitzer} photometry due to blending issues from 2 nearby sources as shown in Figure \ref{fig:slitpos} but note that future \textit{HST}/COS and \textit{JWST} imaging would help spatially resolve and constrain the rest-frame UV and infrared properties of EELG1002. In total, we used 32 photometric bands with 29 drawn from COSMOS2020/Classic and a single \textit{HST}/WFC3 F140W detection drawn from ancillary imaging as described below (\S\ref{sec:WFC3}). The \textit{CFHT}/WIRCam $H$ and $K_s$ photometry was drawn from COSMOS2015 \citep{Laigle2016} due to the lack of UltraVISTA $YJHK_s$ coverage of EELG1002. 
		
		\begin{table}
			\centering
			\caption{Multiwavelength photometry of EELG1002 used in our SED fitting. All data are drawn from the COSMOS2020 Classic \citep{Weaver2022} except for CFHT/WIRCam (COSMOS2015; \citealt{Laigle2016}) and \textit{HST}/WFC3 F140W (\S\ref{sec:WFC3}). EELG1002 has \textit{GALEX} and \textit{Spitzer} detections but we do not include them in our SED fitting due to blending issues with 2 nearby sources.}
			\label{table:photometry}
			\begin{tabular*}{\columnwidth}{@{\extracolsep{\fill}}lccc}
				\hline
				Band & $\lambda_c$ & FWHM & mag$_\textrm{\texttt{AUTO}}$ \\
				          & (\AA) & (\AA) & (AB mag) \\
				\hline
				\multicolumn{4}{l}{\textbf{CFHT/MegaCam}} \\
				$u$ & 3709 & 518 & $23.84\pm0.01$\\			
				$u^*$ & 3858 & 598 & $23.93\pm0.01$\\					
				\multicolumn{4}{l}{\textbf{Subaru/HyperSuprimeCam}} \\	
			    $g$ & 4847 & 1383 & $23.84\pm0.01$\\			
			    $r$ & 6219 & 1547& $23.82\pm0.01$\\			
			    $i$ & 7699 & 1471 & $23.82\pm0.01$\\		
			    $z$ &8894 & 766 & $22.12\pm0.01$\\
			    $y$ & 9761 & 786 & $24.20\pm0.04$\\	
				\multicolumn{4}{l}{\textbf{Subaru/SuprimeCam}} \\						
				$B$ & 4488 & 892 & $24.06\pm0.01$\\
				$g^+$ & 4804 & 1265 & $23.87\pm0.01$\\
				$V$ & 5487 & 954 & $23.78\pm0.02$\\
				$r^+$ & 6305 & 1376 & $23.84\pm0.02$\\
				$i^+$ & 7693 & 1497 & $23.84\pm0.03$\\
				$z^+$ & 8978 & 847 & $22.56\pm0.02$\\
				$z^{++}$ & 9063 &  1335 & $22.54\pm0.01$\\
				$IB427$ & 4266 & 207 & $23.93\pm0.04$\\
				$IB464$ & 4635 & 218 & $23.76\pm0.04$\\
				$IA484$ & 4851 & 229 & $23.99\pm0.04$\\
				$IB505$ & 5064 & 231 & $23.78\pm0.04$\\
				$IA527$ & 5261 & 243 & $23.93\pm0.03$\\
				$IB574$ & 5766 & 273 & $24.00\pm0.05$\\
				$IA624$ & 6232 & 300 & $23.76\pm0.04$\\
				$IA679$ & 6780 & 336 & $23.43\pm0.03$\\
				$IB709$ & 7073 & 316  & $23.68\pm0.03$\\
				$IA738$ & 7361 & 324 & $23.83\pm0.05$\\
				$IA767$ & 7694 & 365 & $24.38\pm0.08$\\
				$IA827$ & 8243 & 343 & $24.31\pm0.07$\\
				$NB711$ & 7121 & 72 & $23.26\pm0.06$\\
				$NB816$ & 8150 & 120 & $24.11\pm0.06$ \\
				\multicolumn{4}{l}{\textbf{\textit{HST}}} \\
				ACS/F814W & 8333 & 2511 & $23.37\pm0.01$\\
				WFC3/F140W & 13923 & 3933 & $23.89\pm0.08$\\
				\multicolumn{4}{l}{\textbf{CFHT/WIRCam}} \\
				$H$ & 16243 & 2911 & $23.82\pm0.26$\\	
				$K_s$ & 21434 & 3270 & $24.39\pm0.44$ \\																																					
				\hline
			\end{tabular*}
		\end{table}
		
		\subsection{\textit{HST}/WFC3 F140W imaging}
		\label{sec:WFC3}
		
		Ancillary \textit{HST}/WFC3 F140W imaging of EELG1002 was found on the MAST portal. The associated program, PID \#15115 (PI: John Silverman; \citealt{Silverman2018,Ding2020}), targeted a sample of $z \sim 1.5$ broad-line AGN to study supermassive black hole properties and how it relates to galaxy mass. One of their main targets, LID360, happened to be oriented such that EELG1002 fell within the \textit{HST} FoV with a total exposure time of $\sim 2400$ seconds. Post-processing and photometry measurements were made as part of the Hubble Advance Program (HAP) Single Visit Mosaics (SVM) program. EELG1002 was found to have a F140W magnitude of $23.89\pm0.08$ mag within a $2''$ aperture (large enough to encompass the full source; see Figure \ref{fig:slitpos}).

		\section{Methodology}
            \label{sec:methodology}
            
		\subsection{Data Reduction}
		We reduce all raw spectroscopic data using \pypeit~(\citealt{Prochaska2020}; \texttt{v1.10dev} -- \texttt{gemini\_gmos\_mask\_ingestion} branch\footnote{At the time data reduction was done, GMOS mask ingestion was only available through the \texttt{v1.10dev} version and \texttt{gemini\_gmos\_mask\_ingestion} branch. This has since been incorporated in newer official release versions of \pypeit.}), a \texttt{Python} semi-automated data reduction pipeline that supports many ground- and space-based facilities. The pipeline starts by reading the setup files that list the location of the science/standard \& associated calibration frames. These setup files are made available via the linked GitHub repository\footnote{\href{https://github.com/akhostov/EELG1002}{https://github.com/akhostov/EELG1002}} for anyone to reproduce the final spectra shown in this work. The first main script of the pipeline automatically does bias \& overscan subtraction, slit edge detections, flat fielding, removal of cosmic rays, wavelength calibrations, tilt corrections, sky subtraction, flexure corrections, and a first pass on object extraction using the \cite{Horne1986} optimal extraction algorithm. 
        
		The second step in the pipeline is to 2D coadd all 3 reduced science frames of varying central wavelengths (accounts for GMOS chip gaps in final 2D coadded spectra). To account for potential drift during observations, we define the offsets using the brightest object in the mask. \pypeit~then models the spatial profile of the object and calculates the spatial shift in reference to the first science frame in our 2D coadd list. This is then stacked with weights corresponding to the S/N of our brightest object. Objects are then extracted using the \cite{Horne1986} optimal extraction algorithm.

		\subsubsection{Flux Calibration \& Slit Loss Corrections}
		
		Flux calibration is performed by first reducing our standard star LTT7379 (G0) using the same \pypeit~reduction as mentioned above. We then used the wavelength-calibrated, reduced spectra to measure the sensitivity curve that converts raw counts into physical flux units. \pypeit~does this by comparing the observed 1D GMOS spectra against archival spectra drawn, in this case, from the ESO Optical and UV Spectrophotometric Standard Stars library. The final sensitivity curve is then applied to each of our science targets. Atmospheric extinction corrections are applied to both the standard star used in generating the sensitivity curve and the science target spectra taking into account the varying airmass at the time of observation. This is all included in the \pypeit~scripts that make the sensitivity curve and apply the flux calibration. This was done by using the observatory's extinction file (in our case, La Silla) and applying the extinction curve as a function of wavelength and air mass associated with the spectra being corrected.
		
		An additional correction is also necessary to account for slit loss (e.g., a fraction of the total flux is lost because the object extends off the slit). Figure \ref{fig:slitpos} shows the configuration of the GMOS slit along with the Subaru / HSC \textit{i} contours demonstrating how a significant fraction of light falls outside the slit. We corrected for this factor using the spatially resolved \textit{HST}/ACS F814W image also shown in Figure \ref{fig:slitpos} and measure the total flux collected along the slit. We then smoothed out the F814W image using a Gaussian kernel with FWHM set to the seeing ($0.9''$) and again measured the total flux within the $0.5''$ slit. We measure 45.6\% of the total flux from EELG1002 is lost and correct the 1D spectra by multiplying a factor of 1.456 to take into account the slit loss.
		
		\subsection{Spectroscopic Redshift \& Emission Line Profile Fitting}
		\label{sec:specz_line_profile}
		
		We used \pyqsofit~\citep{Guo2018} for fitting the profiles of all emission lines observed in the GMOS spectra, which requires an initial estimate of the visually measured spectroscopic redshift ($z = 0.8275$) using \verb|SpecPro| \citep{Masters2011}. Each detected emission line is fit by a single Gaussian profile with the central wavelength allowed to vary to capture variations in the final measured spectroscopic redshift. Line widths of each emission line are set to be constant among other associated lines within the same atomic specie (e.g., FWHMs of the Balmer line are the same from \hbeta~to H$\eta$). This allows us to deblend lines from different atomic species that are in close wavelength proximity to one another (e.g., \neiii3968\AA~and H$\epsilon$ separated by 2.6\AA; rest-frame) and are not fully resolved given the GMOS wavelength resolution limit. The \oii3726,3729\AA~doublet is also not resolved, for which we fit two Gaussians assuming the FWHM from \oiii5007\AA~to fit the doublet. Line fluxes are measured by integrating the best-fit Gaussian profiles. We do not include Balmer absorption corrections as such features were not observed (see Figure \ref{fig:spectra_and_sed}) and are not expected for such a young galaxy as will be shown further in this paper.
		
		
		Uncertainty measurements are measured by sampling the posterior probability distributions associated with each parameter in the emission line profiles via Markov Chain Monte Carlo (MCMC) using \texttt{emcee} \citep{Foreman2013}. \pyqsofit~does this by using the best-fit parameters of each line profile as the initial starting point in the chain perturbed by a small random offset and assuming 100 walkers. To ensure convergence and well-sampling of the true posterior probability distributions, \texttt{emcee} keeps track of the autocorrelation time which measures how many steps are required to ensure independent samples are used with the condition that the number of samples is greater than 50 times the autocorrelation time.  After several tests, we determined 15000 samples with a 20\% burn-in (discard the initial steps) satisfies this criteria. Overall, each emission line detected has an associated probability distribution that will be used throughout this work to measure line ratios and associated uncertainties.
	
		\subsection{Velocity Dispersion}
		\label{sec:velocity_dispersion}
		
		The observed velocity dispersion, $\sigma_{obs}$, is measured using the best-fit Gaussian emission line profile. We measure the `intrinsic' velocity dispersion, $\sigma_{int}$, by:
		\begin{equation}
			\centering
			\sigma_{int} = \sqrt{ \sigma_{obs}^2 - \sigma_{inst}^2 - \sigma_{th}^2 }
		\end{equation}
		which subtracts out the effects of instrumental ($\sigma_{inst}$) and thermal ($\sigma_{th}$) broadening. The latter is measured as $\sigma_{th} = \sqrt{k_B T_e / m_X}$, where $k_B$ is the Boltzmann constant, $m_X$ is the mass of atomic species (e.g., H, O, Ne), and $T_e$ is the electron temperature for which we use our measurement described in \S\ref{sec:ne_and_te} and Table \ref{table:ISM}. The instrumental broadening is measured using the FWHM of the CuAr arc lines ($2.5$ pix with $3.88$ \AA~pix$^{-1}$ resolution).

		\subsection{Measuring ISM Properties}
		\label{sec:ism_methods}
		In this section, we will describe how we use nebular emission line ratios in conjunction with \verb|PyNeb| \citep{Luridiana2015} to measure key ISM properties. We present the key line ratios with the shorthand notation used in this work and its exact definitions (e.g., lines used) in Table \ref{table:ratio_defs}.
		
		\begin{table}
			\centering
			\caption{Key Line Ratio shorthands and definitions used throughout this paper.}
			\label{table:ratio_defs}
			{\renewcommand{\arraystretch}{1.1}
				\begin{tabular*}{\columnwidth}{@{\extracolsep{\fill}}ll}
					\hline
					ID & Definition \\
					\hline
					O3HB & $\textrm{\oiii}5007\textrm{\AA}/\textrm{\hbeta}$\\
					O32 & $ \textrm{\oiii}5007\textrm{\AA}/\textrm{\oii}3726,3729\textrm{\AA}$\\
					Ne3O2 & $ \textrm{\neiii}3869\textrm{\AA}/\textrm{\oii}3726,3729\textrm{\AA}$\\
					Ne3O3 & $ \textrm{\neiii}3869\textrm{\AA}/\textrm{\oiii}5007\textrm{\AA}$\\
					$R2$ & $\textrm{\oii}/\textrm{\hbeta}$\\
					$R3$ &$(\textrm{\oiii}5007\textrm{\AA} + \textrm{\oiii}4959\textrm{\AA})/\textrm{\hbeta}$\\
					$R23$ & $(\textrm{\oiii}5007\textrm{\AA} + \textrm{\oiii}4959\textrm{\AA} + \textrm{\oii})/\textrm{\hbeta}$\\
					\hline
				\end{tabular*}
			}
		\end{table}

		\subsubsection{Dust Extinction}
		
		Balmer decrements (e.g., \halpha/\hbeta) trace the level of dust extinction within H{\sc ii} regions and are used to dust-correct observed line fluxes when coupled with an assumed dust attenuation curve. The reddening is measured by comparing the observed-to-intrinsic Balmer ratios as shown below:
		\begin{eqnarray}
			E(B-V) = \frac{2.5}{k(\textrm{\hgamma}) - k(\textrm{\hbeta})} \log_{10} \Bigg(\frac{ [\textrm{\hbeta}/\textrm{\hgamma}]_\textrm{\scriptsize obs} }{[\textrm{\hbeta}/\textrm{\hgamma}]_\textrm{\scriptsize int}}\Bigg)
		\end{eqnarray}
		coupled with the dust attenuation curve evaluated at the wavelength of each Balmer line used, $k(\lambda)$, for which we assume the \citet{Calzetti2000} attenuation curve. The amount of dust extinction is then measured as $A_\lambda = k(\lambda) E(B-V)$ at any given wavelength. The GMOS spectral coverage does not include \halpha~for which we instead use the next brightest ratio, \hbeta/\hgamma, to measure $E(B-V)$ and $A_\lambda$. The intrinsic line ratio, \hbeta/\hgamma~$ = 2.11$, is determined using \pyneb~assuming case B recombination, $T_e = 15000$ K (electron temperature) and $n_e = 100$ cm$^{-3}$ (electron density). Varying these assumptions marginally changes the intrinsic ratios and does not affect our dust-extinction measurements. Uncertainties in $E(B-V)$ are measured by each MCMC realization used in the fitting of the emission line profile (see \S\ref{sec:specz_line_profile}).
		
		\subsubsection{Electron Temperature \& Density}
		\label{sec:ne_and_te}
		The electron temperature is measured using a two-zone H{\sc ii} region model due to \oiii~(doubly ionized oxygen) tracing the (high ionizing) zone closest to the ionizing source and \oii~(single ionized oxygen) being able to trace the outer (low ionizing) zone. Electron temperature of the high ionizing zone, $T_e(\rm{O}^{++})$, is measured using the auroral \oiii4363/\oiii5007\AA~line ratio at a fixed $n_e = 1000$ cm$^{-3}$ and the \texttt{getTemDen} function in \pyneb. The fixed $n_e$ is done arbitrarily as $T_e$ is essentially independent of electron density up to $\sim 10^4$ cm$^{-3}$ (not expected for ISMs of galaxy populations) but is an input in \texttt{getTemDen}.
		
		Measuring the low ionizing zone electron temperature requires O$^+$ auroral lines (e.g., \oii7322,7332\AA) which are outside our GMOS coverage. Instead, we estimate O$^+$ electron temperature using the \citet{Izotov2006} calibration:
		\begin{eqnarray}
			\centering
			t_e(\textrm{O}^+) = t_e(\textrm{O}^{++}) \times (2.065 - 0.498 t_e(\textrm{O}^{++}))  - 0.577
		\end{eqnarray}
		with $T_e  = t_e\times 10^4$ K for both O$^+$ and O$^{++}$. This assumes the low metallicity condition described in \citet{Izotov2006} which is found to be appropriate for EELG1002 given its metal-poor nature (see \S\ref{sec:metallicity} and Table \ref{table:ISM}).

		We measure the electron density, $n_e$, using the \oii3726,3729\AA~doublet which is sensitive to $T_e$ (e.g., can vary from $\sim 500$ cm$^{-3}$ to $\sim 640$ cm$^{-3}$ by increasing $T_e$ from 10000 K to 20000 K assuming the \oii~doublet ratio is of order unity.) Therefore, we first measure the electron temperature as described above and then measure $n_e$ using \texttt{getTemDen} function with  $T_e(\rm{O}^{+})$ and observed \oii3729/\oii3726\AA~ratio.

		\subsubsection{Gas-Phase Abundance}
		\label{sec:abundances}
		Abundances associated with Oxygen and Neon are measured using \pyneb~and its \texttt{getIonAbundance} feature assuming the electron temperatures in both the high and low ionizing zones. Our approach assumes $n_e$ is constant throughout the ionizing zones. Below we describe how the ionic and total abundances are measured.
		
		Oxygen abundance is measured with the formalism:
		\begin{eqnarray}
			\frac{\textrm{O}}{\textrm{H}} \approx \frac{\textrm{O}^+}{\textrm{H}}  + \frac{\textrm{O}^{++}}{\textrm{H}} 
			\label{eqn:oxygen_abundance}
		\end{eqnarray}
		where additional abundance contribution from O$^{+++}$ is not expected to contribute significantly to the overall O/H abundance. Specifically, to have O$^{+++}$ requires $54.9$ eV and, therefore, we would also expect Helium recombination lines (e.g., He{\sc i}4686\AA) for which we find no detections in the GMOS spectra. \cite{Berg2018} investigated a $z \sim 2$ lensed source with O{\sc iv} detection and found that O$^{+++}/$H contribution only accounted for a $\sim 5$\% increase ($+0.02$ dex) in the total O/H abundance. Therefore, we continue measuring O/H abundance as defined in Equation \ref{eqn:oxygen_abundance}. The $\textrm{O}^{++}/\textrm{H}$ abundance is measured using $R3$ with $n_e$ and $T_e(\rm{O}^{++})$ measured in \S\ref{sec:ne_and_te} and $\textrm{O}^{+}/\textrm{H}$ is measured using $R2$, $n_e$, and $T_e(\rm{O}^{+})$. 

		The neon abundance is defined as:
			\begin{eqnarray}
				\frac{\textrm{Ne}}{\textrm{H}} \approx \frac{\textrm{Ne}^{++}}{\textrm{H}} \times \textrm{ICF}
			\end{eqnarray}	
		where the ionization correction fraction (ICF) traces the relative contribution from other Neon ionization states (e.g., Ne$^+$). We adopt the ICF formalism of \citet{Izotov2006}: 
		\begin{equation}
				\textrm{ICF} = 1.365 -0.385\frac{\textrm{O}^{++}}{\textrm{O}^{+} + \textrm{O}^{++}} + 0.022\frac{\textrm{O}^{+} + \textrm{O}^{++}}{\textrm{O}^{++}}
		\end{equation}		
		which assumes a low metallicity system (consistent with EELG1002; see Table \ref{table:ISM}). The Ne$^{++}$/H abundance is measured using the \neiii3869\AA/\hbeta~ratio along with our measured $n_e$ and $T_e(\rm{O}^{++})$ as \neiii~requires 41 eV comparable to the 35 eV needed for \oiii~emission. Although EELG1002 has strong \neiii3968\AA~emission (see Figure \ref{fig:spectra_and_sed}), we do not include it in the abundance measurement due to uncertainties from H$\epsilon$ blending. 

		\subsubsection{Ionization Parameter}
		
		The ionization parameter is defined as:
		\begin{equation}
			U = \frac{Q_{H}}{4 \pi  R_S^2 c n_H}
			\label{eqn:logU}
		\end{equation}
		where $Q_H$ is the number of hydrogen ionizing photons per unit time, $R_S$ is the Str\"{o}mgen radius, $c$ is the speed of light, and $n_H$ is the hydrogen number density (comparable to the electron density, $n_e$). Essentially, $U$ traces how many ionizing photons are produced within an H{\sc ii} region in comparison to the present number of hydrogen atoms. A higher ionization parameter would signify more ionizing photons and/or fewer hydrogen atoms that can be interpreted as energetic ISM conditions.
		
		O32 is a classical indicator of $U$ as O$^{++}$ requires 35 eV compared to O$^{+}$ which requires 13.6 eV such that it is sensitive to the ionization state of gas within H{\sc ii} region. An alternative is the Ne3O2 ratio where Ne$^{++}$ requires $\sim 41$ eV and is insensitive to dust attenuation \citep{Levesque2014}. Coupling of both can also provide insight as to the shape (hardness) of the ionizing spectrum given Ne$^{++}$ has ionization potential $\sim 6$ eV higher that O$^{++}$ (e.g., \citealt{Strom2017,Jeong2020}). Using both diagnostics to directly measure $U$ is limited based on the gas-phase metallicity range used in calibrating the diagnostic \citep{Kewley2019}. As such, we measure $U$ via photoionization modeling included in \texttt{Bagpipes} \citep{Carnall2018} and \texttt{Cigale} \citep{Boquien2019}.
		
		In making measurements of $U$, one must be cautious about electron densities assumed given its inverse proportionality. The typical assumption of $100$ cm$^{-3}$ can overestimate (underestimate) $U$ if the source of interest has $n_e > 100$ ($< 100$) cm$^{-3}$. We take this into account in our spectrophotometric SED fitting (\S\ref{sec:SED}) by using $n_e = 1000$ cm$^{-3}$ in the older \texttt{Cloudy} model within \texttt{Cigale} (pre-defined grid; \S\ref{sec:cigale}) and recompute the \texttt{Cloudy} grid used in \bagpipes~(\S\ref{sec:bagpipes}) assuming $n_e = 800$ cm$^{-3}$ consistent with EELG1002 (Table \ref{table:ISM}).

		\subsection{Spectral Energy Distribution Fitting}
		\label{sec:SED}
		
		In this section, we describe our SED fitting parameters and assumptions used in both \texttt{Cigale} and \texttt{Bagpipes} where we simultaneously fit both the GMOS spectra and ancillary multi-wavelength photometry (see \S\ref{sec:data}). Although \textit{GALEX}/FUV and NUV photometry do exist or EELG1002, we chose to disregard it in the SED fitting process given the low spatial resolution as shown in Figure \ref{fig:slitpos} where blending with 2 nearby sources is an issue. The same is true for \textit{Spitzer}/IRAC photometry. Lastly, all photometry used in the SED fitting process has an additional 10\% uncertainty included in quadrature to account for underestimated errors. Setup files are also made available within the \texttt{GitHub} repository \footnote{\href{https://github.com/akhostov/EELG1002}{https://github.com/akhostov/EELG1002}}.

		\subsubsection{{\sc Cigale}}
		\label{sec:cigale}
				
		We use \texttt{Cigale v2022.1} \citep{Boquien2019,Yang2022}, a template grid-based SED fitting code, and provide a brief overview of assumptions made in the SED fitting process which are highlighted in Table \ref{table:cigale}. Both photometry and line fluxes measured from the GMOS spectra are included in the SED fitting process. Briefly, we assume a \cite{Bruzual2003} stellar population synthesis model coupled with a \cite{Chabrier2003} IMF and consider stellar metallicities within the range of 0.005 and 1 $Z_\odot$. Nebular emission is modeled using the \cite{Inoue2011} templates which were generated using \texttt{Cloudy v13.01} \citep{Ferland2013} with $U$ considered in the range of $-3.0$ to $-1.0$. Gas-phase metallicity is fixed to 0.001 (closest to our best metallicity measurement; Table \ref{table:ISM}). \texttt{Cigale} only allows for three different fixed values for $n_e$ for which we assume $1000$ cm$^{-3}$ given our measured $n_e$ (see Table \ref{table:ISM} and \S\ref{sec:ne_and_te}). Both $f_{esc} $ (LyC escape fraction) and $f_{dust}$ (fraction of LyC photons absorbed by dust) are fixed to $0$\% in the nebular emission modeling; however, we do note that there is the possibility of a nonzero $f_{esc}$ given \textit{GALEX}/FUV detection (blended with 2 nearby sources) and we discuss the prospects of LyC escape for EELG1002 in \S\ref{sec:fesc}. All emission lines in the model have fixed observed FWHM set to $400$ km s$^{-1}$, consistent with our observations (Table \ref{table:line_fits}). $E(B-V)$ is fixed to 0 mag given \hbeta/\hgamma~(see Table \ref{table:ISM}).

		We assume a delayed-$\tau$ star formation history with a recent burst. The main/old stellar population is considered with e-folding time, $\tau_{main}$, of 50 Myr to 7.5 Gyr and ages of 50 Myr to 6.6 Gyr. Setting $\tau_{main} > 6.6$ Gyr (age of Universe at $z = 0.8275$) would describe a continuously rising SFH. The young stellar population formed in the recent burst is modeled with e-folding time, $\tau_{burst}$, between 1 and 100 Myr with ages between 1 to 50 Myr. The contribution of stellar mass formed in the recent burst towards the stellar mass of the galaxy, $f_{burst}$, is considered between 0 and 99\%, where the latter is the maximum allowed in \texttt{Cigale} and is interpreted as a galaxy for which the total stellar mass was formed recently in a single burst of star formation.

		\subsubsection{Bagpipes}
		\label{sec:bagpipes}
		
		We use \texttt{Bagpipes} \citep{Carnall2018}, a Bayesian spectrophotometric SED fitting suite, to take advantage of several features: nonparametric star formation history (SFH) modeling, inclusion of binary stellar populations, updated photoionization models from \texttt{Cloudy v17} \citep{Ferland2017}, and direct fitting to the 1D spectra (\texttt{Cigale} only fits to provided line fluxes). \texttt{Bagpipes} uses \texttt{MultiNest} \citep{Feroz2008,Feroz2009,Feroz2019}, a nested sampling package that samples the parameter space given a likelihood, and bayesian inference to measure best-fit properties. Table \ref{table:bagpipes} highlights our defined parameter space and all setup files are available via our \texttt{GitHub} repository. Briefly, we use \texttt{BPASS v2.2.1} \citep{Stanway2018} which includes binary stellar populations with an assumed broken power law IMF with a slope of $-1.35$ (0.1 -- 0.5 \msol) and $-2.35$ (0.5 -- 300 \msol). The shallower slope at low stellar masses allows for an increased contribution of the older, low-mass stellar population compared to \citet{Chabrier2003} IMF (exponentially cuts off at 0.1 -- 1 \msol). Nebular metallicity is fixed to our gas-phase metallicity shown in Table \ref{table:ISM} while the stellar metallicity is a free parameter. Reddening is also set to 0 mag given our measured Balmer Decrement suggests no dust extinction. Templates for the nebular component are recomputed with \texttt{Cloudy v17.03} using the default \texttt{BPASS} stellar grid provided in \texttt{Bagpipes} to cover $\log_{10} U = -4$ to $-1$ in order to consider cases of extreme energetic ISM conditions. To ensure reliable ionization parameter measurements, we recompute the \texttt{Cloudy} grid to also assume $n_e = 800$ cm$^{-3}$ consistent with our source (see Table \ref{table:ISM}).
		
		The non-parametric SFH is modeled using the \cite{Leja2019} continuity formalism within \texttt{Bagpipes} which separates the stellar mass contribution within inputted time bins. We follow the time bin spacing used in past non-parametric SFH modeling studies (e.g., \citealt{Leja2017,Leja2019,Tacchella2022,Tang2022}): 0, 3, 10, 30, 100, 300, 1000, 3000, and 6000 Myr. These are logarithmically spaced equally ($\sim 0.5$ dex) except for the 3000 -- 6000 Myr bin which is limited by the age of the Universe. Shorter time bins gauge how rapidly the recent burst of star formation occurred and are constrained primarily by the multiple strong Balmer line detections tracing instantaneous star formation activity while longer time bins factor in past star formation that formed the older stellar population.

	\subsection{Measuring Equivalent Width}
	\label{sec:measure_ew}
	Equivalent width (EW) is measured as the ratio between emission line flux (\S\ref{sec:specz_line_profile}) and its associated continuum flux density (\S\ref{sec:SED}). The latter is measured using the best-fit \texttt{Cigale} and \texttt{Bagpipes} SEDs by first masking out the emission line of interest. We then select continuum flux densities within 2 windows: one bluewards and the other redwards of the emission line. Both windows are 10\AA~(rest) in width and are placed $\pm10$\AA~(rest) from the emission line center. We place the windows for \oiii5007\AA~and \oiii4959 at 15\AA~away from the line center and with a width of 20\AA~given how strong the lines are (e.g., ensuring we are not probing the line profile wings in the continuum flux density measurement).  In the case of \hgamma, we set the red window to be 35\AA~from center (10\AA~width) to ensure we do not include the nearby \oiii4363\AA~line for which that line also has its blue window placed closer ($-6$\AA~from line center and extending to $-15$\AA). \neiii3869~also has the red window placed 35\AA~from center (width 10\AA) to ensure H8 emission is not included in the window. The same is true for H8 with the blue window set to $-35$\AA~from line center.
	
	Using the continuum flux densities measured within each window, we interpolate the SED shape between both windows and measure the flux density about the emission line center. We then take the ratio of the emission line flux and the determined continuum flux density and measure the equivalent width of the line which are shown in Table \ref{table:line_fits} and labeled based on whether the best-fit \texttt{Cigale} or \texttt{Bagpipes} SED was used in determining the continuum flux density.
	
	\subsection{Ionizing Photon Production Efficiency}
	\label{sec:xi_ion}
	
	The ionizing photon production efficiency, \xiion, traces how well a galaxy can produce ionizing photons (e.g., $> 13.6$ eV that can ionize H{\sc i}) and is defined as:
	\begin{eqnarray}
		\textrm{\xiion} = \frac{Q_H}{L_{UV}^{int}}
		\label{eqn:xiion}
	\end{eqnarray}
	where $Q_H$ represents the production rate of hydrogen ionizing photons and $L_{UV}^{int}$ is the intrinsic, rest-frame, dust-corrected $1500$\AA~continuum luminosity. The definition of $\xi_{ion}$ is slightly varied where we refer the reader to \S3.2 of \cite{Chevallard2018} for a detailed discussion. One definition typically cited as $\xi_{ion}^\star$ uses the monochromatic UV luminosity attributed only to the stellar continuum and ignores the emission and absorption caused by neutral and ionized gas. Another definition cited as \xiionHtwo~incorporates the nebular and stellar continuum in measuring the $1500$\AA~monochromatic luminosity ($L_{UV}^\textrm{H{\sc ii}}$). We adopt this definition throughout this paper and use both \xiionHtwo~and \xiion~interchangeably unless otherwise clarified if we are referring to another definition (e.g., $\xi_{ion}^\star$).
	
	We use the best-fit combined stellar and nebular continuum fits from \texttt{Cigale} and \texttt{Bagpipes} to measure $L_{UV}$ using a tophat filter centered at $1500$\AA~with a width of 100\AA. We use the \cite{Leitherer1995} calibration:
	\begin{eqnarray}
		Q_H = 2.1\times10^{12} L_{int}(\textrm{H}\beta)~\textrm{[s$^{-1}$]}
	\end{eqnarray}
	to measure the ionizing photon production rate, $Q_H$, with \hbeta~luminosity from our observed GMOS spectra. This assumes a Lyman Continuum escape fraction $f_{esc} = 0$ and represent the maximum $Q_H$ possible based on the calibration. Given that Balmer Decrements show $E(B-V) \sim 0$ mag, we do not apply any dust correction.

	\subsection{Galaxy Size \& Dynamical Mass Measurement}
	\label{sec:dynamical_mass}
	
	We use archival \textit{HST} ACS/F814W to measure the rest-frame $\sim 4500$\AA~galaxy size. A single Sersic profile is assumed and fitted to the F814W image using \texttt{pysersic} \citep{Pasha2023}: a public package that uses Bayesian inference to fit Sersic profiles to galaxy images. We also assume a flat background around EELG1002 which is simultaneously fit during the Sersic profile fitting. Assumed priors are the default in \texttt{pysersic}. The effective/half-light radius, $r_{e}$, has a truncated normal prior centered at 3.52 pixels with $\sigma$ of 3.75 pixels and a minimum cutoff at 0.5 pixels. The Sersic index, $n$, is set to a uniform prior between 0.65 and 8. The posterior for both variables is measured using a No U-turn Sampler (NUTS, \citealt{Hoffman2014}) and is used to measure the uncertainties.
	
	The dynamical mass is measured based on the half-light radius and intrinsic velocity dispersion (\S\ref{sec:velocity_dispersion}) and traces all baryonic matter within the galaxy: stellar, gas, dust, and dark matter. Given that we have measurements of stellar mass from our SED fitting and that EELG1002 has relatively no dust (Table \ref{table:ISM}), the dynamical mass compared to the stellar mass provides a tracer of how much gas and dark matter resides within EELG1002. The dynamical mass is defined as:
	\begin{equation}
		M_{dyn} = C \frac{\sigma_{int}^2 r_e}{G}
		\label{eqn:dynamical_mass}
	\end{equation}
    where $r_e$ is the effective radius, $G$ is the gravitational constant, $\sigma_{int}$ is the intrinsic velocity dispersion (corrected for instrumental and thermal broadening; see \S\ref{sec:velocity_dispersion}), and $C$ is the scaling factor. We use the \hbeta~$\sigma_{int}$ as shown in Table \ref{table:line_fits}. The scaling factor is dependent on the mass distribution and velocity field of the galaxy and could range from $C \sim 1$ -- 5 \citep{Erb2006}. We follow \citet{Maseda2013} which measured dynamical masses for EELG populations at $z \sim 2$ and assumed $C = 3$. However, we also incorporate the range of $C \sim 1$ -- $5$ in our uncertainties by uniformly sampling $C$ between $1$ and $5$ and measuring the dynamical mass. This provides for a conservative estimate on the possible range in dynamical mass.

		\section{Results}
            \label{sec:results}
		\newcommand{\ISMandLineTable}{\input{ISM_and_Line_properties.table}}
		\begin{table}
		\centering
		\caption{Measured ISM properties of EELG1002 with the methodology described in \S\ref{sec:ism_methods}.}
		\label{table:ISM}
		{\renewcommand{\arraystretch}{1.12}
			\begin{tabular*}{\columnwidth}{@{\extracolsep{\fill}}lc}
				\hline
				ISM Property & Measurement \\
				\hline
				\input{ISM_and_Line_properties.table}
				\hline
				\end{tabular*}
			}
		\end{table}

		\begin{table*}
			\centering
			\caption{Emission Line Properties for all detected lines in the GMOS spectra. Line flux and velocity dispersions, $\sigma$, are measured using \pyqsofit~with FWHM fixed based on \hbeta, \oiii, and \neiii~for the Hydrogen, Oxygen, and Neon atomic species, respectively. The spectroscopic redshifts are also based on the brightest line of each species and is found to be consistent with $z \sim 0.8275$ within $1\sigma$. Rest-frame Equivalent Width (EW$_0$) measurements use the best-fit SED (\cigale~and \bagpipes) for measuring the continuum flux density about the emission line wavelength. Observed velocity dispersion, $\sigma_{obs}$, is based on the best-fit FWHM from the emission line profile. Intrinsic velocity dispersion, $\sigma_{int}$, corrects $\sigma_{obs}$ for instrumental and thermal broadening (\S\ref{sec:velocity_dispersion}).}
			\begin{tabular*}{\textwidth}{@{\extracolsep{\fill}}lcccccc}
				\hline
				Line & $z_{spec}$ & Observed Line Flux  & EW$_0$ (\verb|Cigale|)& EW$_0$ (\verb|Bagpipes|) & $ \sigma_{obs}$& $\sigma_{int}$\\
				&                     &  ($10^{-17}$ erg s$^{-1}$ cm$^{-2}$) & ($10^{-17}$ erg s$^{-1}$ cm$^{-2}$) & (\AA) & (km s$^{-1}$)  & (km s$^{-1}$) \\
				\hline
				\multicolumn{7}{l}{{\textbf{Balmer Lines}}}\\
				\hbeta & $0.8273^{+0.0005}_{-0.0005}$ & $18.74^{+0.87}_{-0.87}$ & $472.84^{+22.04}_{-22.03}$ & $403.64^{+18.81}_{-18.81}$ & $192^{+7}_{-6}$ & $131^{+9}_{-10}$ \\
				H$\gamma$ & --- & $9.65^{+0.49}_{-0.48}$ & $187.12^{+9.41}_{-9.26}$ & $157.02^{+7.90}_{-7.77}$ & --- & --- \\
				H$\delta$ & --- & $4.06^{+0.32}_{-0.32}$ & $71.93^{+5.65}_{-5.64}$ & $57.31^{+4.50}_{-4.49}$ & --- & --- \\
				H$\epsilon$ & --- & $2.26^{+0.38}_{-0.38}$ & $36.84^{+6.24}_{-6.23}$ & $29.49^{+5.00}_{-4.99}$ & --- & --- \\
				H$\zeta+$He{\sc i}3889\AA & --- & $0.71^{+0.33}_{-0.32}$ & $10.50^{+4.88}_{-4.72}$ & $8.41^{+3.91}_{-3.78}$ & --- & --- \\
				H$\eta$ & --- & $3.02^{+0.31}_{-0.31}$ & $46.83^{+4.74}_{-4.80}$ & $37.36^{+3.78}_{-3.83}$ & --- & --- \\
				\hline
				\multicolumn{7}{l}{{\textbf{Oxygen Lines}}}				\\    	    			    					
				\oiii5007\AA & $0.8276^{+0.0005}_{-0.0005}$ & $92.15^{+0.63}_{-0.64}$ & $2438.63^{+16.75}_{-17.05}$ & $2039.37^{+14.01}_{-14.25}$ & $145^{+1}_{-1}$ & $52^{+3}_{-3}$ \\
				\oiii4959\AA & --- & $30.43^{+0.39}_{-0.41}$ & $786.47^{+10.16}_{-10.69}$ & $658.25^{+8.50}_{-8.95}$ & --- & --- \\
				\oiii4363\AA & --- & $2.97^{+0.40}_{-0.40}$ & $59.93^{+8.01}_{-8.08}$ & $48.02^{+6.42}_{-6.47}$ & --- & --- \\
				\oii3726,3729\AA & --- & $10.28^{+0.46}_{-0.46}$ & $154.89^{+6.88}_{-6.88}$ & $126.57^{+5.63}_{-5.62}$ & --- & --- \\
				\oii3726\AA & --- & $5.30^{+0.74}_{-0.75}$ & $79.11^{+11.03}_{-11.20}$ & $67.83^{+9.46}_{-9.60}$ & --- & --- \\
				\oii3729\AA & --- & $4.98^{+0.73}_{-0.73}$ & $68.75^{+10.11}_{-10.08}$ & $59.09^{+8.69}_{-8.67}$ & --- & --- \\
				\hline
				\multicolumn{7}{l}{{\textbf{Neon Lines}}}		\\		    	    			    						    		    
				\neiii3869\AA & $0.8275^{+0.0006}_{-0.0006}$ & $11.06^{+0.49}_{-0.47}$ & $159.79^{+7.01}_{-6.84}$ & $131.48^{+5.77}_{-5.63}$ & $232^{+12}_{-11}$ & $141^{+19}_{-20}$ \\
				\neiii3968\AA & --- & $3.42^{+0.00}_{-0.00}$ & $55.54^{+0.00}_{-0.00}$ & $44.65^{+0.00}_{-0.00}$ & --- & --- \\
				\hline
			\end{tabular*}
			\label{table:line_fits}
		\end{table*}

		\subsection{Record Breaking High EW, Low Mass Extreme Emission Line Galaxy at $z \sim 0.8$}
		
		\begin{figure*}
			\centering
			\includegraphics[width=\textwidth]{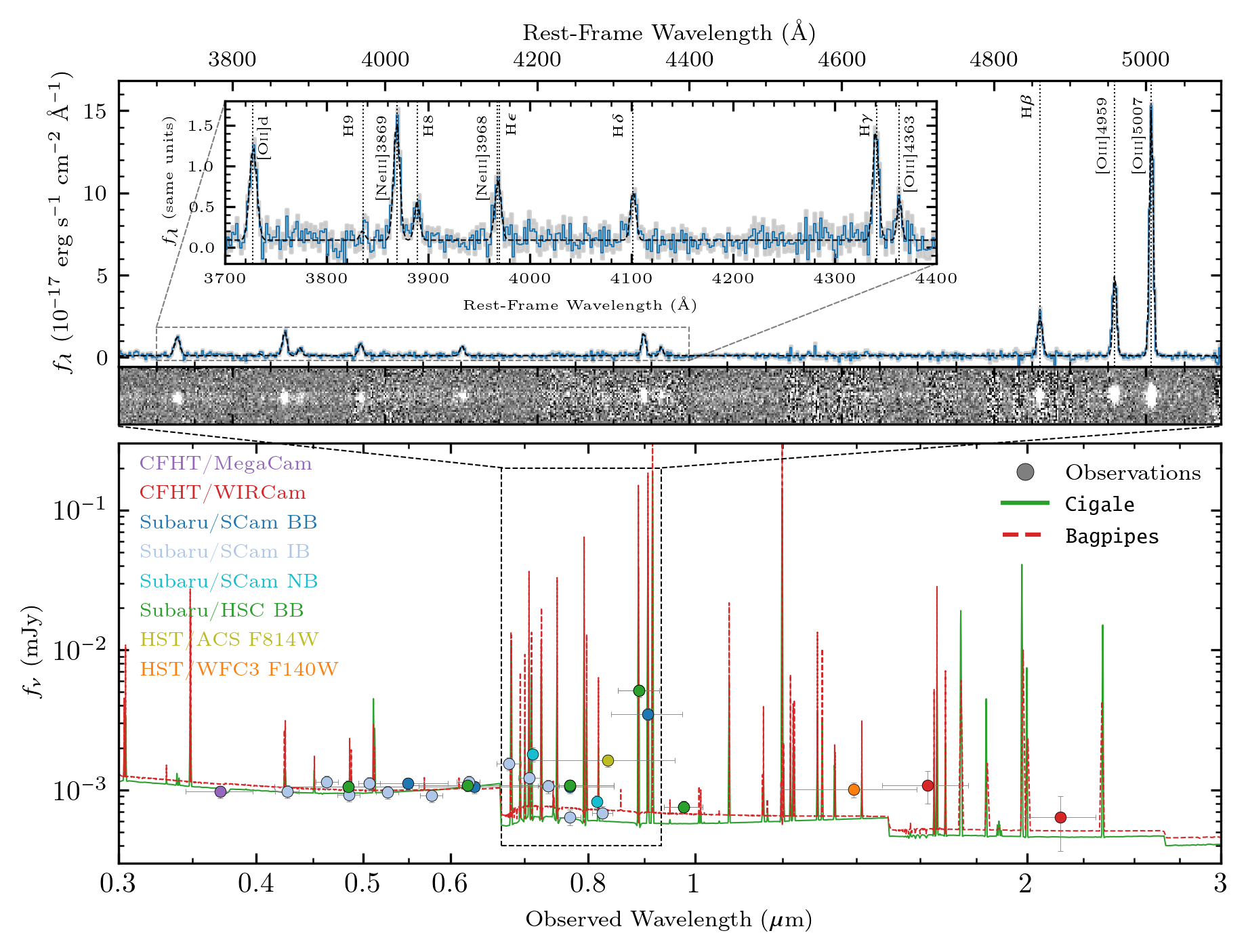}
			\caption{\textit{Top:} Reduced GMOS spectra of EELG1002 highlighting multiple emission line detections securing the spectroscopic redshift with the fitted emission line spectrum (\textit{black dashed line}). Strong \oiii5007\AA~relative to \oii3727\AA~emission along with comparable \neiii3869\AA~and \oii~emission highlight the highly energetic ISM. \oii4363\AA~detection allows for direct-$T_e$ abundance measurements confirming the low-metallicity nature. \textit{Bottom:} Multiwavelength photometry and best-fit SEDs for EELG1002 with \cigale~(\textit{green}) and \bagpipes~(\textit{red}). Strong \oiii+\hbeta~color excess is clearly observed in Subaru/HSC and SCam $i$ indicative of high EWs. \textit{HST} F814W also shows an excess although not as sensitive given the wider wavelength coverage. Subaru intermediate and narrowbands are also affected by strong nebular emission line features. \cigale~and \bagpipes~are in relative agreement with slight differences. We note stellar population, IMF, and SFH modeling assumptions may be driving these minor differences.}
			\label{fig:spectra_and_sed}
		\end{figure*}
	
		\begin{figure}
			\centering
			\includegraphics[width=\columnwidth]{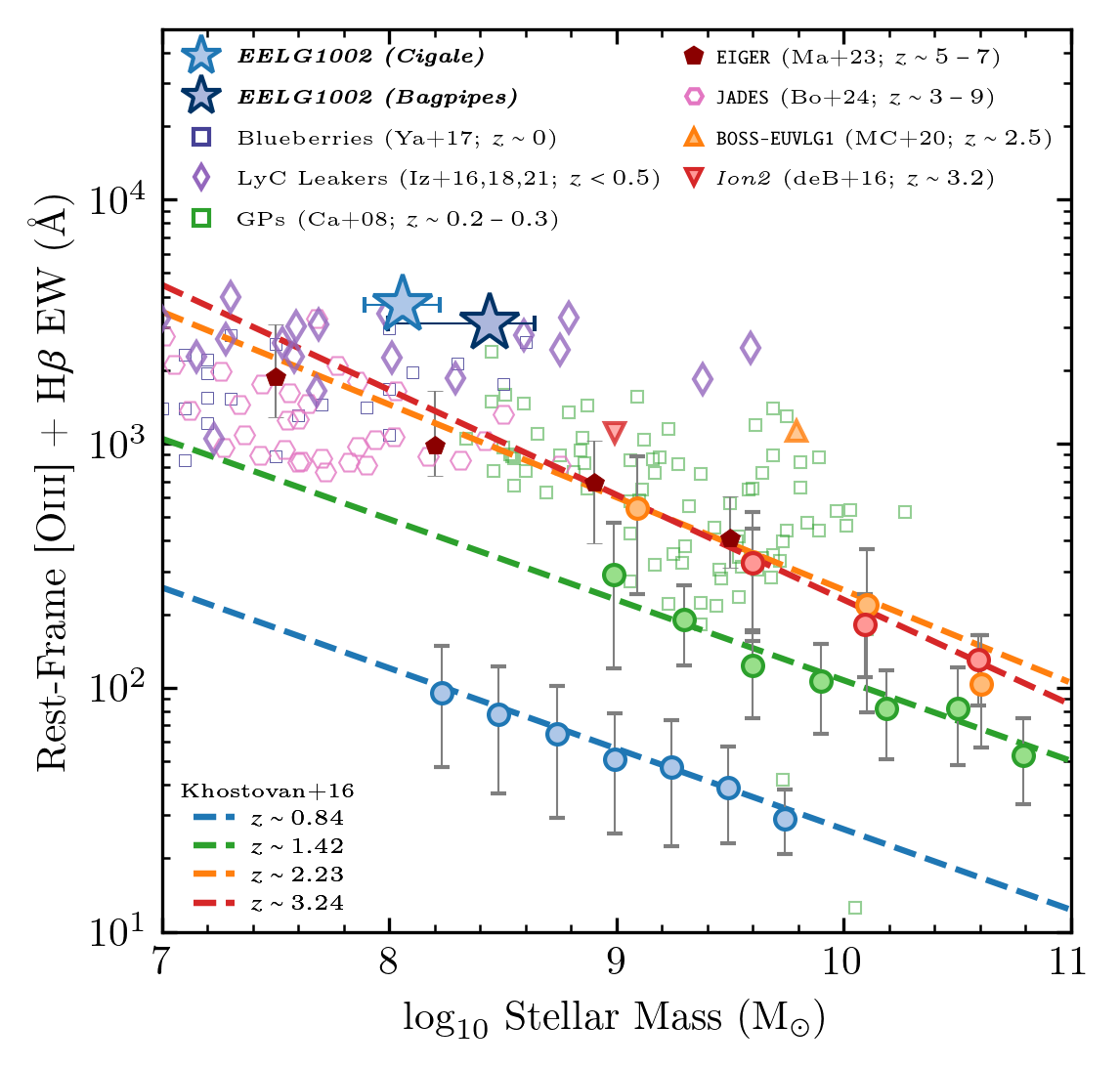}
			\caption{The rest-frame \oiii+\hbeta~EW -- stellar mass anti-correlation. EELG1002 is shown as a \textit{star} with \textit{dark blue} using \bagpipes~and \textit{light blue} for \cigale~for the continuum flux density in measuring \oiii+\hbeta~EW. We find EELG1002 is a uniquely `extreme' ELG with \oiii+\hbeta~EW $\sim 32$ -- $36\times$ higher compared to the typical \oiii$+$\hbeta~EW at $z \sim 0.8$ (\textit{blue line}; \citealt{Khostovan2016}) and somewhat higher than the typical EW at $z \sim 3$ -- $9$ \citep{Khostovan2016, Matthee2023, Boyett2024}. EELG1002 also has EW higher than Blueberries \citep{Yang2017_BB}, Green Peas \citep{Cardamone2009}, the $z \sim 2.5$ intense starburst BOSS-EUVLG1 \citep{Marques-Chaves2020}, and the $z \sim 3.2$ LyC emitter \textit{Ion2} \citep{deBarros2016}. EELG1002 is consistent with local LyC leakers \citep{Izotov2016,Izotov2018,Izotov2021} and we do discuss the potential of LyC escape further in this study. Overall, EELG1002 is a uniquely rare and `extreme' ELG with EWs somewhat more extreme than EoR-era galaxies.}
			\label{fig:EW}
		\end{figure}
		EELG1002 ($\alpha = 10:00:32.304$, $\delta = +2:51:11.351$) is a uniquely low-mass, high EW emission line galaxy with numerous emission line detections, as shown in the top panel of Figure \ref{fig:spectra_and_sed}, that, in conjunction with the wealth of multi-wavelength ancillary photometry, provides us a great deal of information in regards to its star-formation and ISM conditions. We present redshift, line flux, EW, and velocity dispersion measurements for each line in Table \ref{table:line_fits}. We find a spectroscopic redshift in the range of $z_{spec} \sim 0.8273$ and $0.8276$ depending on which line is used (\hbeta, \oiii5007\AA, and \neiii3869\AA). The discrepancies are well within $1\sigma$ error bars and could arise simply due to uncertainties within the wavelength calibration and limiting resolution of the GMOS R150 grating. Throughout this work, we quote the spectroscopic redshift as $z_{spec} \sim 0.8275$ which is the average of the three different measurements. 
		
		Figure \ref{fig:spectra_and_sed} shows all the multi-wavelength photometry associated with EELG1002 and our best-fit \cigale~and \bagpipes~SEDs. The photometry alone shows the presence of nebular emission line contribution in the narrowband, intermediate band, and broadband photometries (e.g., Subaru SuprimeCam and HSC $z$-band is dominated by the presence of \oiii+\hbeta) which would suggest high EW emission lines. \cigale~(\textit{green}) and \bagpipes~(\textit{red}) SEDs are for the most part consistent in showing strong nebular emission line features, but vary in several key areas. Both \cigale~and \bagpipes~show a strong inverse Balmer jump and slightly increasing UV slope consistent with the presence of young stellar populations.
		
		We find EELG1002 has a stellar mass of $(2.75^{+1.61}_{-1.77}) \times 10^8$ \msol~(\bagpipes) and $(1.14\pm0.44) \times 10^8$ \msol~(\cigale). Using the continuum fluxes as described in \S\ref{sec:measure_ew}, we find EELG1002 has significantly strong EWs as shown in Table \ref{table:line_fits}. The combined rest-frame \oiii+\hbeta~EW for EELG1002 is $3101^{+25}_{-25}$\AA~(\bagpipes) and $3697^{+30}_{-30}$\AA~(\cigale) which is significantly high for a $z \sim 0.8$ emission line galaxy. To understand how `extreme' EELG1002 is requires that we take this within the context of how elevated the \oiii+\hbeta~EW is in comparison to typical star-forming galaxies at the same stellar mass and redshift. Figure \ref{fig:EW} shows the typical EWs of emission line galaxies at a given stellar mass and redshift from $z \sim 0.8$ to $3.2$ measured from narrowband-selected \oiii~emitters \citep{Khostovan2016}. We find that EELG1002 lies 32 (\bagpipes) to 36 (\cigale) times above the typical \oiii+\hbeta~EWs at $z \sim 0.8$ placing this well above the typical range of EWs of star-forming galaxies at similar redshift and stellar mass.
		
		In both \cigale- and \bagpipes-measured continuum and stellar mass, EELG1002 is also `extreme' relative to typical $z \sim 3.2$ \citep{Khostovan2016} and even $z \sim 3$ -- 7 \oiii~emitters observed recently with \textit{JWST} via EIGER \citep{Matthee2023}  and JADES \citep{Boyett2024}. EELG1002 is also consistent in terms of its EW with $z \sim 0$ confirmed $z< 0.5$ LyC Leakers (e.g., \citealt{Izotov2016,Izotov2018,Izotov2021}) and Blueberries \citep{Yang2017_BB} (compact, high EW, extreme emission line galaxies at $z \sim 0$) and is even more `extreme' in its \oiii+\hbeta~EW compared to known high-$z$ LyC leakers. For example, \textit{Ion2} \citep{deBarros2016} has \oiii+\hbeta~EW $\sim 1103$\AA~which is $\sim 2.2\times$ higher than typical $z \sim 3$ \oiii~emitters. BOSS-EUVLG1 \citep{Marques-Chaves2020} has \oiii+\hbeta~EW $\sim 1125$\AA~which is $\sim 3.7\times$ higher than typical $z \sim 2.2$ EWs.
		
		Overall, EELG1002 has $\sim 32 - 36 \times$ higher \oiii+\hbeta~EW compared to the typical EW of a star-forming galaxy at similar stellar mass and redshift making it a potentially record-breaking system in this regard. It also has EWs at or larger than the typical EWs currently observed by high-$z$ studies with \textit{JWST} making it uniquely placed as an analog of high-$z$ galaxies and provides an opportunity given the wealth of emission lines detected in the GMOS spectra to investigate star-formation and ISM properties in great detail.

		\subsection{No Evidence of AGN component}
		\label{sec:AGN}
		
		\begin{figure}
			\centering
			\includegraphics[width=\columnwidth]{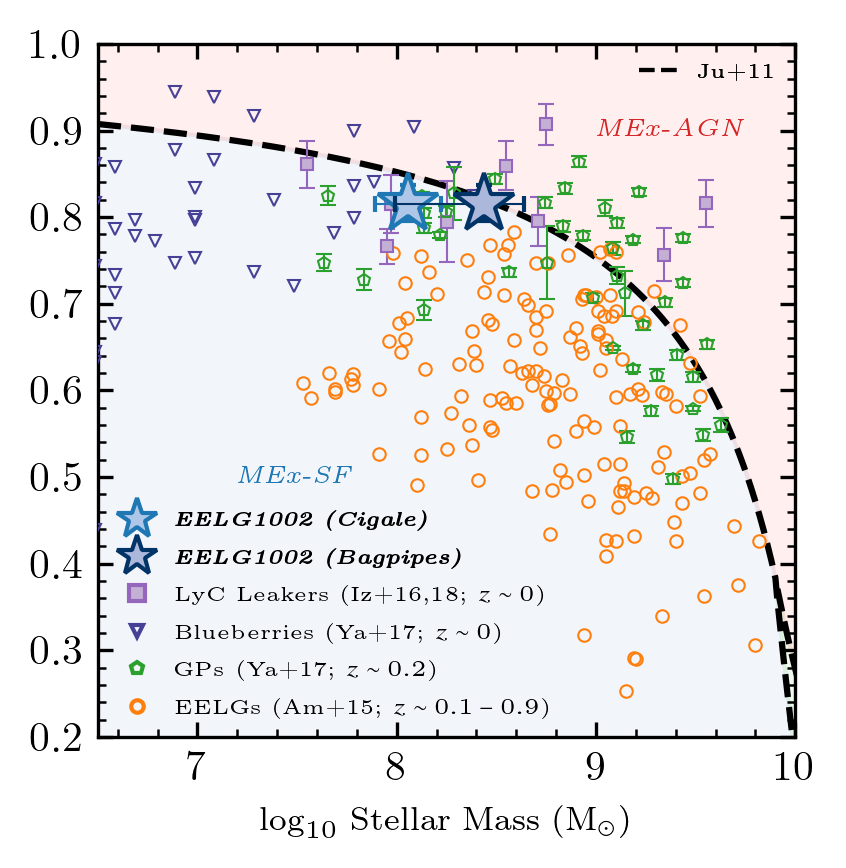}
			\caption{Mass Excitation (MEx) diagram for SFG/AGN classification where EELG1002 is found to have \oiii/\hbeta~ratios and stellar mass consistent with star-forming galaxies. \bagpipes-derived stellar mass is near the boundary; however, this can also be due to the different IMF assumed. \oiii/\hbeta~ratios in conjunction with its low direct $T_e$ measured metallicity is consistent with star-forming galaxies with low \nii/\halpha~in the BPT diagram \citep{Pettini2004, Marino2013}. EELG1002 also has \oiii/\hbeta~ratios consistent with Blueberries \citep{Yang2017_BB} and Green Peas \citep{Yang2017_GP}. Based on this diagnostic, EELG1002 shows no evidence of an AGN component.}
			\label{fig:BPT_MEx}
		\end{figure}

		Although EELG1002 has uniquely high \oiii+\hbeta~EW given its redshift and stellar mass, it does raise the question: \textit{is such strong nebular emission driven by an active galactic nuclei (AGN)? }We address this question and possibility using all the available evidence that we have on hand.
		
		We first investigate the GMOS spectra for signatures of broad line emission that could be an indication of Type 1 AGN (`BL-AGN', FWHM $> 1000$ km s$^{-1}$; e.g., \citealt{Stirpe1990,Ho1997}). However, as shown in Figure \ref{fig:spectra_and_sed}, we find no evidence for a broad line component in any of the observed emission lines. Furthermore, the observed velocity dispersion reported in Table \ref{table:line_fits} range from 145 to 232 km s$^{-1}$ and correcting for instrumental resolution and thermal broadening reduces the velocity dispersion to between 52 and 141 km s$^{-1}$ which is well below the $> 1000$ km s$^{-1}$ FWHM typically observed in BL-AGN (e.g., \citealt{Genzel2014}).
		
		We also consider the possibility of a potentially strong ionizing radiation spectrum (see \S\ref{sec:ionization}) that could produce strong ionization lines such as [Ne{\sc v}]3426\AA~(requires $\sim 94$ eV) indicative of an AGN. However, we find no evidence within the GMOS spectra for the presence of [Ne{\sc v}] emission. We also inspect \textit{Chandra}-COSMOS \citep{Civano2016,Marchesi2016} and \textit{XMM} imaging (PI: G. Hasinger; e.g., \citealt{Hasinger2007}) and find no hard and soft X-ray detections.
		
		We lastly use the star-forming/AGN classification diagnostic -- Mass-Excitation diagram (MEx; \citealt{Juneau2011}). MEx is very similar to BPT \citep{BPT} where instead of \nii/\halpha, the diagnostic uses the mass-metallicity relation traced via \nii/\halpha~to convert the BPT into a $R3$ -- stellar mass diagnostic. Figure \ref{fig:BPT_MEx} shows that EELG1002 lies entirely within the star-forming galaxy classification region for the \cigale-measured stellar mass which assumes the same \citet{Chabrier2003} IMF used in the MEx diagram. \bagpipes-measured stellar mass also falls within the star-forming galaxy classification but the upper stellar mass uncertainty does place it slightly within the AGN classification region; however different IMF assumptions between \bagpipes~and the MEx classification must be considered as well. Furthermore, \oiii/\hbeta~ratio in conjunction with the low direct $T_e$ metallicity (e.g., low \nii/\halpha) is consistent with metal-poor star-forming galaxies within the classic BPT diagram.
			
		Given the available data, we conclude that there is no evidence for the presence of AGN activity within EELG1002 and that the emission lines are most likely driven by star-formation processes. This does not necessarily mean that an AGN is not fully present within EELG1002. One potential possibility is an IR AGN; however, due to the blending of \textit{Spitzer}/IRAC photometry and the lack of spatially resolved \textit{JWST} NIRCam and MIRI imaging, we can not assess if there is a dusty AGN in EELG1002. Overall, we find no evidence based on the available data that would support an AGN component although future infrared imaging could shed light on the possibility of a dusty AGN component.

		\subsection{Extremely Metal-Poor, Chemically Unevolved System Consistent with $z \gtrsim 5$ Galaxies}
		\label{sec:metallicity}
		\begin{figure}
			\centering
			\includegraphics[width=\columnwidth]{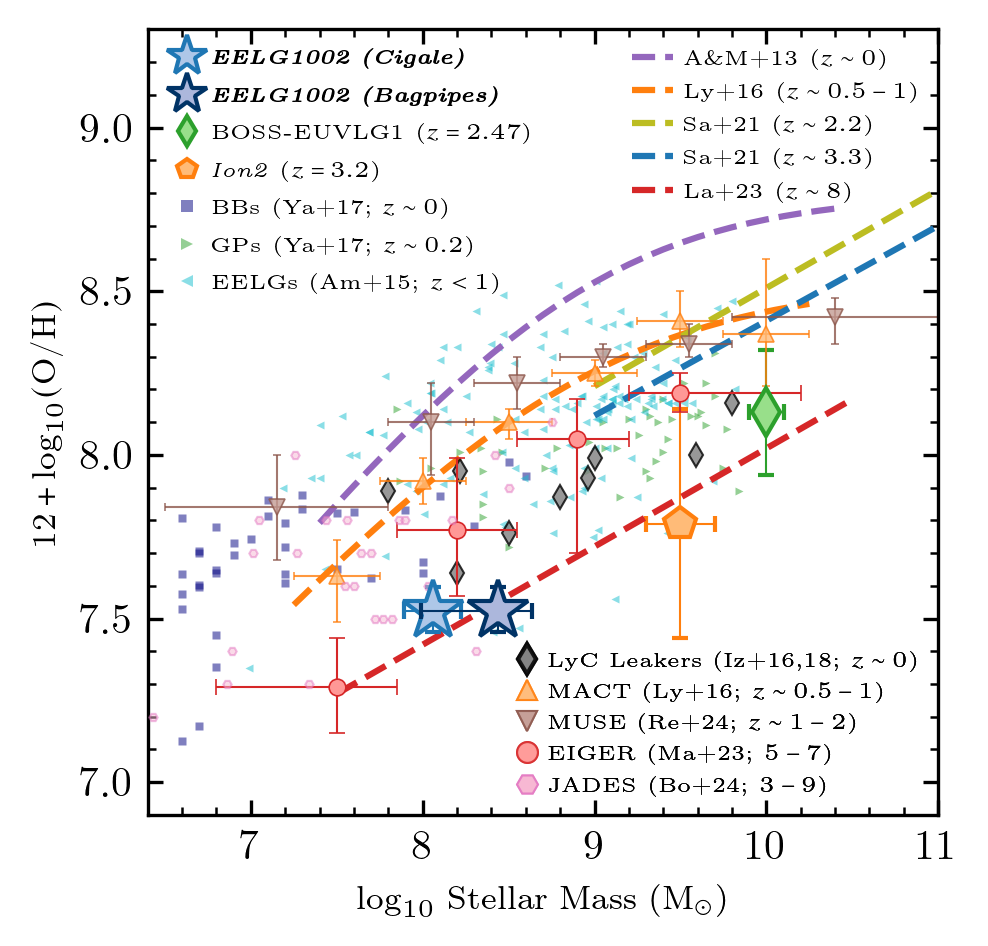}
			\caption{Mass -- Metallicity Relation (MZR) with measured relations at $z \sim 0$ \citep{Andrews2013}, $z \sim 0.5 - 1$ \citep{Ly2016}, $ z\sim 2.2$ -- $3.3$ \citep{Sanders2021}, and $z \sim 8$ \citep{Langeroodi2023}. EELG1002 is found to have $12+\log_{10}$ (O/H) below $z < 1$ EELGs \citep{Amorin2015}, Blueberries \citep{Yang2017_BB}, and Green Peas \citep{Yang2017_GP} at fixed stellar mass. Both \cigale~and \bagpipes~stellar mass measurement places EELG1002 highly consistent with $z \sim 5$ -- $9$ galaxies at similar stellar mass \citep{Matthee2023, Langeroodi2023, Boyett2024}. Similar to EELG1002, Local LyC leakers \citep{Izotov2016,Izotov2018}, \textit{Ion2} \citep{deBarros2016}, and BOSS-EUVLG1 \citep{Marques-Chaves2020} are consistent with high-$z$ MZRs but with higher stellar mass and $12 + \log_{10}$ (O/H) suggesting past star formation and chemical enrichment periods. EELG1002 is found to have both low-mass and low metallicity consistent with high-$z$ MZR suggesting a chemically-unevolved system that could be undergoing a first bursty phase of star-formation activity as expected of high-$z$ galaxies (e.g., \citealt{Cohn2018}).}
			\label{fig:MZR}
		\end{figure}

		One key feature of the GMOS spectra is the clear $\sim 7.4\sigma$ detection of the auroral \oiii4363\AA~emission line which enables direct electron temperature (\S\ref{sec:ne_and_te}) and oxygen abundance measurements (\S\ref{sec:abundances}). We find EELG 1002 is quite metal poor with  12+$\log_{10}$(O/H) $ = 7.52\pm0.07$ ($Z_{gas} = 0.068^{+0.013}_{-0.009} Z_\odot$ assuming solar gas-phase metallicity ($12+\log_{10}$(O/H)$ = 8.69$; \citealt{Asplund2021}). This is also consistent with the $R23$ -- O$32$ calibration of \cite{Jiang2019} which is based on direct $T_e$ measurements using strong emission line emitters where we find 12+$\log_{10}$(O/H) $ \sim 7.57$. 
  
        In comparison to measured mass-metallicity relations (MZRs), we show in Figure \ref{fig:MZR} that EELG1002 is well below the $z \sim 0$ \citep{Andrews2013}, $z \sim 0.5$ -- 1 ($\mathcal{MACT}$, \citealt{Ly2016}), and $z \sim 2$ -- 3 \citep{Sanders2021} MZRs. EELG1002 is also lower in metallicity compared to known analogs of high-$z$ galaxies such as Blueberries \citep{Yang2017_BB}, Green Peas \citep{Yang2017_GP}, and $0.1 < z < 1$ EELGs \citep{Amorin2015}.
		
		Using both \cigale- and \bagpipes-measured stellar mass, we find EELG1002 is in strong agreement with the $z\sim8$ MZR \citep{Langeroodi2023}, $5 < z < 7$ EIGER \citep{Matthee2023} and $3 < z < 9$ JADES measurements \citep{Boyett2024}. In comparison to known local LyC leakers \citep{Izotov2016,Izotov2018}, EELG1002 is found to be slightly metal-poor at similar stellar mass. However, both EELG1002 and local LyC leakers are consistent with both EIGER and \citet{Langeroodi2023} MZRs. Furthermore, the $z \sim 3.2$ LyC leaker \textit{Ion2} is also consistent with the $z \sim 8$ MZR although at higher stellar mass and metallicities. The same is found for the intense starburst BOSS-EUVLG1 \citep{Marques-Chaves2020} which could signify some chemical enrichment has already occurred after a period of intense star-formation activity leading to higher stellar mass and metallicities.
		
		However, EELG1002 is significantly lower in stellar mass and gas-phase metallicity in comparison to \textit{Ion2} and BOSS-EUVLG1 and may be similar to the chemically unevolved conditions expected of galaxies in the early Universe. EELG1002, as will be discussed further in \S\ref{sec:sfh}, is also undergoing an intense period of star formation activity with \hbeta-measured SFR of $7.7\pm0.4$ M$_\odot$ yr$^{-1}$ with a mass doubling time scale of $\sim 15$ -- 35 Myr (see Table \ref{table:SED}). This would suggest conditions where star-formation activity recently occurred but did not have time yet to chemically enrich its ISM. In that regard, EELG1002 is potentially an analog of what conditions were like within the $z > 6$ galaxies.

			\begin{figure}
				\centering
				\includegraphics[width=\columnwidth]{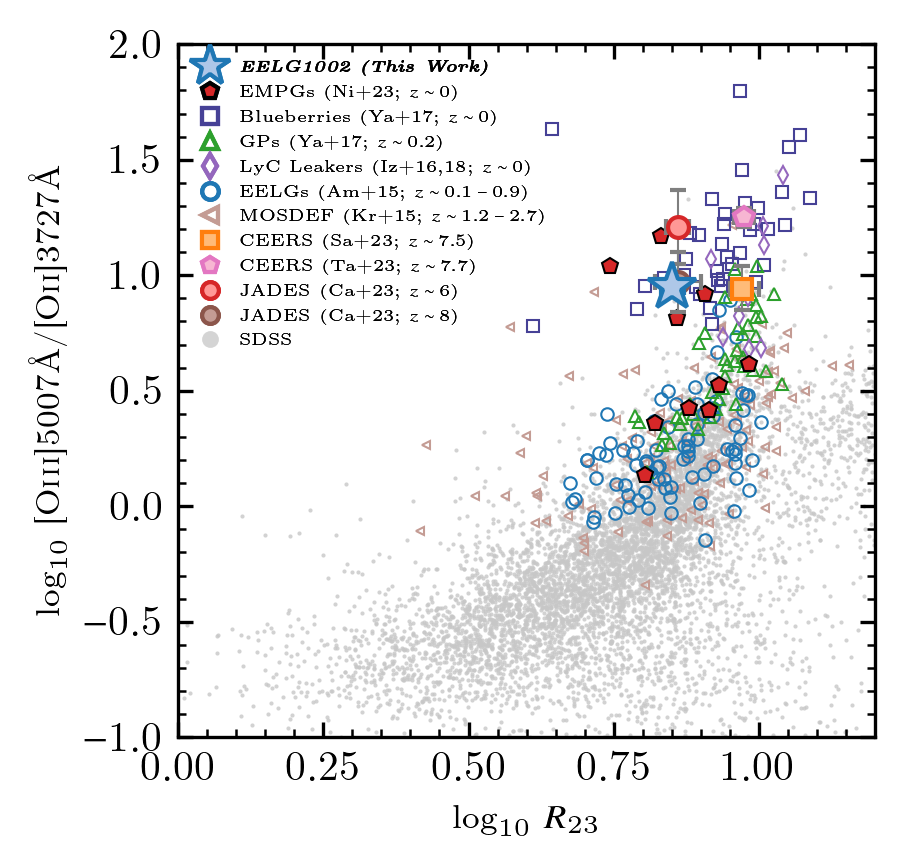}
				\caption{Ionization and Excitation properties. EELG1002 is found to have O32 and $R23$ well above SDSS \citep{Thomas2013}, EELGs \citep{Amorin2015}, Green Peas \citep{Yang2017_GP}, and $z \sim 2$ star-forming galaxies (MOSDEF; \citealt{Kriek2015}). We find EELG1002 has line ratios most consistent with few EMPGs \citep{Nishigaki2023}, Blueberries \citep{Yang2017_BB}, and high-$z$ stacks from CEERS \citep{Sanders2023,Tang2023} and JADES \citep{Cameron2023}. This highlights how EELG1002 has ionization and excitation conditions consistent with EoR-era galaxies making EELG1002 a unique analog in this regard.}
				\label{fig:o32_r23}
			\end{figure}
		
		\subsection{Energetic ISM Conditions \& Harder Ionizing Radiation Field}
		\label{sec:ionization}
		
		As shown in the \textit{top} panel of Figure \ref{fig:spectra_and_sed}, EELG1002 has strong \oiii5007\AA~emission with $\sim (9.2 \pm0.6) \times 10^{-16}$ \cgsline~in line flux and moderate \oii~emission (Table \ref{table:line_fits}) with a measured O32 ratio of $8.92^{+0.44}_{-0.44}$ indicative of energetic conditions (see \S\ref{sec:ionization}). Based on our spectrophotometric SED fitting described in \S\ref{sec:SED}, we find ionization parameters of $\log_{10} U = -2.23 \pm 0.06$ (\cigale) and $-1.96^{+0.06}_{-0.06}$ (\bagpipes) as shown in Table \ref{table:ISM}. The discrepancy is most likely attributed to differences in assumed $n_e$ (default in \cigale~was 1000 cm$^{-3}$ versus recomputed \texttt{Cloudy} grid in \bagpipes~assuming $n_e = 800$ cm$^{-3}$) and the inclusion of binary populations in \bagpipes~via \texttt{BPASS} where young hot, massive stars live for longer periods of time allowing for harder ionizing spectra.
		
		Qualitatively, both \cigale~and \bagpipes~suggests EELG1002 has ionization parameters that would suggest highly energetic conditions within the ISM. The energetic ionization state is aided by the low metallicity conditions described in \S\ref{sec:metallicity} where essentially the number of metal coolants is quite low allowing for higher gas temperatures as we found using \oiii4363\AA~($T_e \sim 19500$ K; Table \ref{table:ISM}). EELG1002 also has $n_e = 779^{+927}_{-487}$ cm$^{-3}$ which is higher than typical electron densities at similar redshift ($\sim 20 - 100$ cm$^{-3}$; e.g., \citealt{Kaasinen2017,Swinbank2019, Davies2021}) and is consistent with $z > 5$ galaxies ($> 200$ cm$^{-3}$; e.g., \citealt{Isobe2023}). Based on Equation \ref{eqn:logU}, this would suggest for elevated levels of hydrogen ionizing photons. 
		
		Figure \ref{fig:o32_r23} shows the O32 (tracer of ionization parameter) and R23 (indcator of gas-phase metallicity) ratios for EELG1002 along with measurements including SDSS DR12 \citep{Thomas2013}, $z \sim 0$ Extremely Metal Poor Galaxies (EMPGs; \citealt{Nishigaki2023}), Blueberries \citep{Yang2017_BB}, Green Peas \citep{Yang2017_GP}, local LyC leakers \citep{Izotov2016,Izotov2018}, EELGs \citep{Amorin2015}, $z \sim 1$ -- 3 MOSDEF \citep{Kriek2015}, $z \sim 7.5$ CEERS \citep{Sanders2023,Tang2023}, and $z \sim 6$ -- 8 JADES \citep{Cameron2023}. We find EELG1002 has O32 and R23 ratios consistent with local Blueberries, local LyC leakers, as well as several EMPGs. It is also consistent with stack measurements from JADES at $z > 6$ highlighting how EELG1002 has both ionization and excitation properties consistent with typical star-forming galaxies in the high-$z$ Universe. 
		
\begin{figure}
			\centering
			\includegraphics[width=\columnwidth]{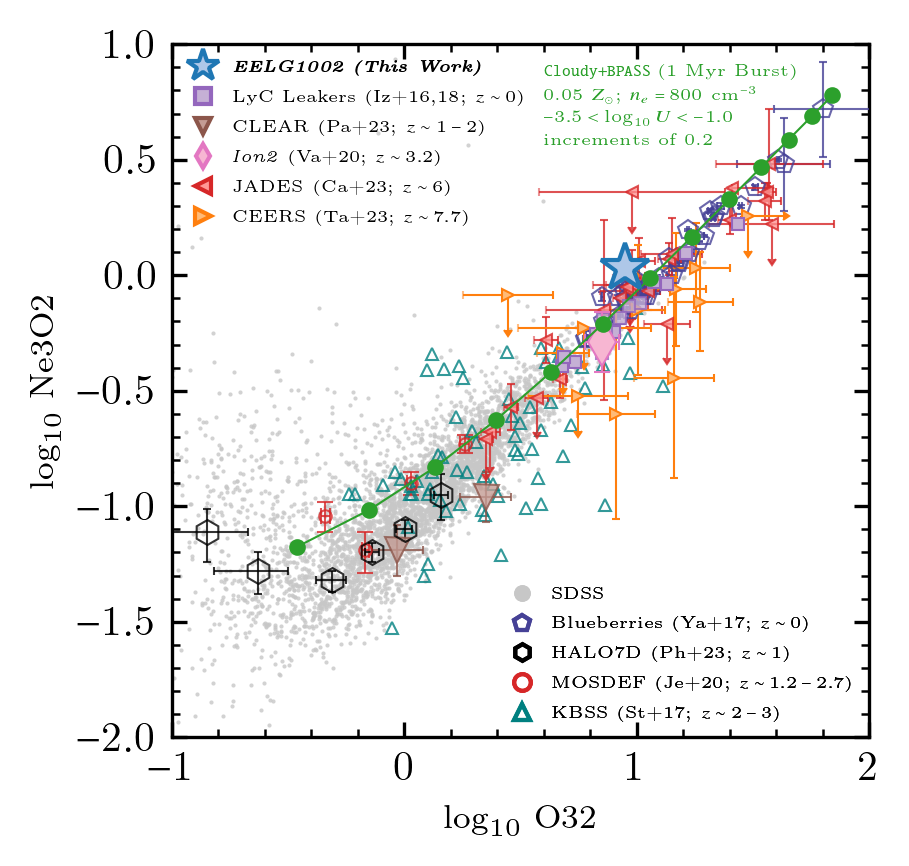}
			\caption{Comparison of Ne3O2 and O32 line ratios as a tracer of harder ionizing radiation spectrum. We find EELG1002 has O32 and Ne3O2 ratios significantly higher than the typical ratios measured by SDSS \citep{Thomas2013}, HALO7D \citep{Pharo2023}, CLEAR \citep{Papovich2023}, MOSDEF \citep{Jeong2020}, and KBSS \citep{Strom2017}. EELG1002 is comparable to known LyC leakers \citep{Izotov2016,Izotov2018},  \textit{Ion2} \citep{Vanzella2020}, and $z>6$ galaxies \citep{Cameron2023,Tang2023}. Compared to our \texttt{Cloudy+BPASS} model (\textit{green}), we find EELG1002 has 0.15 dex higher Ne3O2 at fixed O32 which would suggest a harder ionizing radiation field (e.g., more energetic EUV photons).}
			\label{fig:Ne3O2}
		\end{figure}

		Given the high O32, we can conclude from Figure \ref{fig:o32_r23} that EELG1002 has a highly energetic ISM. However, O32 can also be influenced by the incident radiation field shape. \citet{Sanders2016} showed using photoionization modeling that O32 increases at fixed $U$ with harder ionizing radiation field modeled as a simple blackbody spectrum (e.g., presence of young, hot massive stars). In the case of EELG1002, we can use the Ne3O2 line ratio to gauge the shape of the ionizing radiation field as Ne$^{++}$ requires an ionization potential of 41 eV and, therefore, traces a higher energy regime compared to O$^{++}$ ($\sim 35$ eV). Figure \ref{fig:Ne3O2} shows the Ne3O2 versus O32 line ratios for EELG1002 along with measurements including $z \sim 1 - 2$ CLEAR \citep{Papovich2023}, $z \sim 1$ HALO7D \citep{Pharo2023}, $z \sim 1 - 3$ MOSDEF \citep{Jeong2020}, and $z \sim 2 - 3$ KBSS \citep{Strom2017}. We find EELG1002 has Ne3O2 and O32 line ratios well above the typical $z \sim 1$ galaxies (e.g., \citealt{Pharo2023}) and is more consistent with both high-$z$ galaxies (e.g., \citealt{Cameron2023,Tang2023}) and local `extreme' systems (e.g., \citealt{Izotov2016,Izotov2018,Yang2017_BB}). Figure \ref{fig:Ne3O2} also shows that EELG1002 has somewhat elevated Ne3O2 ratios at fixed O32 in comparison to both typical high-$z$ and $z \sim 0$ `extreme systems. This would suggest the shape of the ionizing radiation field is such that it includes an excess of highly ionizing photons at $> 41$ eV which is enough for Ne$^{++}$. 
		
		To better gauge the potential of a harder ionizing radiation field, we use \texttt{Cloudy} photoionization modeling and \texttt{BPASS} binary stellar population with age of 1 Myr as the incident radiation field. The electron density is assumed to be $n_e = 800$ cm$^{-3}$ with metallicity of $0.05~Z_\odot$ consistent with EELG1002. We allow for varying measurements of $\log_{10} U$ ranging from $-3.5$ to $-1$ in $0.2$ dex increments. Our model is overlaid in Figure \ref{fig:Ne3O2} (\textit{green}) and we find EELG1002 has Ne3O2 ratio $\sim 0.15$ dex (5.5$\sigma$) above the \texttt{Cloudy+BPASS} prediction at fixed O32. As such, we conclude that EELG1002 most likely has an energetic ISM with a hard ionizing radiation field which could also be enhanced given the combination of low gas-phase metallicity (low metal coolants), high electron temperature, and elevated $n_e$ all resulting in elevated levels of ionizing photons.

		\begin{figure*}
			\centering
			\includegraphics[width=\textwidth]{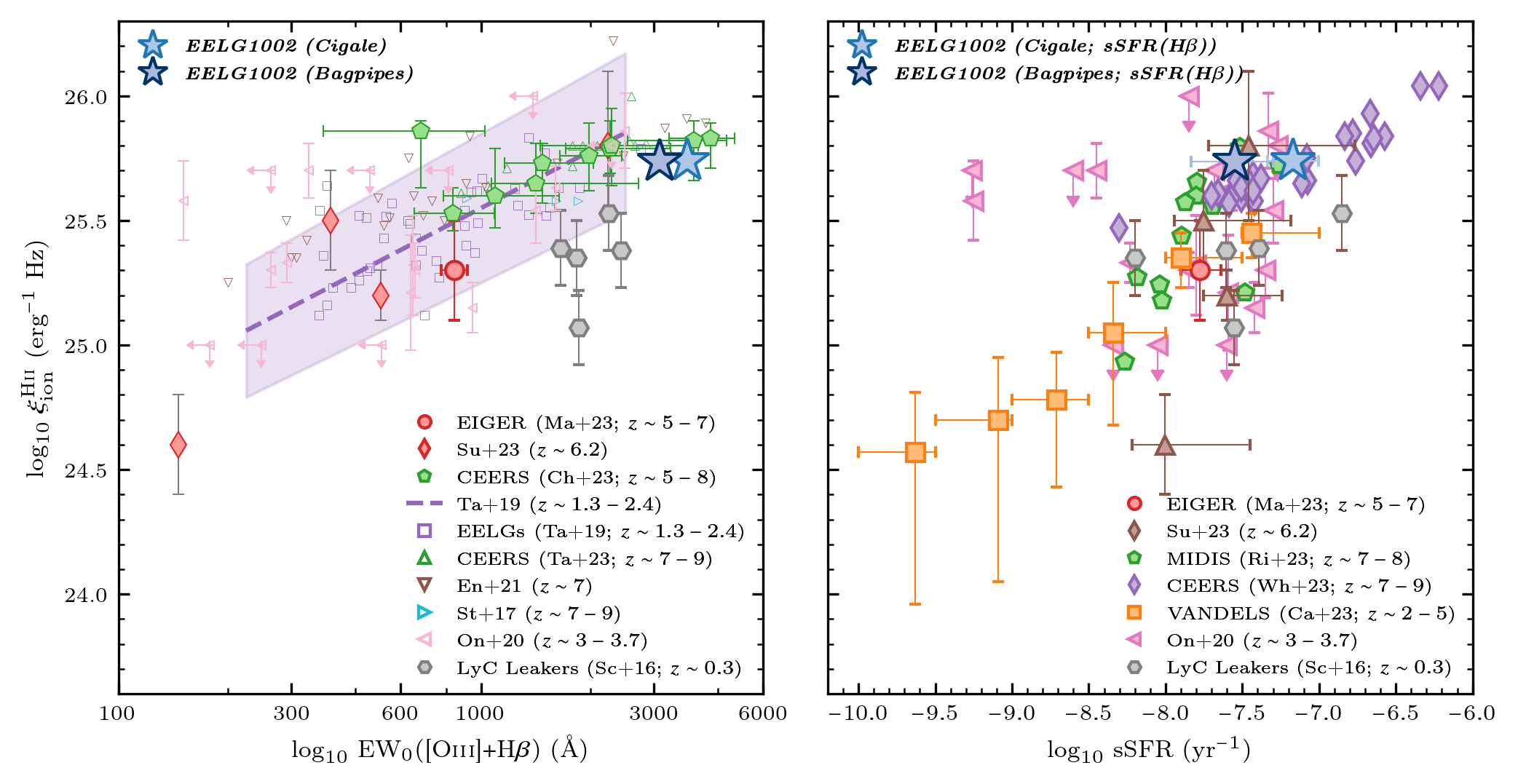}
			\caption{\textit{Left:} Ionizing photon production efficiency, \xiion, in terms of \oiii+\hbeta~EW. EELG1002 is found to have \xiion~higher than $z \sim 1.3$ -- $2.4$ EELGs \citep{Tang2019}, $5 < z < 7$ EIGER \citep{Matthee2023}, and $z \sim 6.2$ \textit{JWST}/ERS \citep{Sun2023}. We find EELG1002 has \xiion~mostly consistent with the highest EW emitters identified at $3 < z < 9$ \citep{Stark2017,Onodera2020,Endsley2021,Tang2023,Chen2024}. \textit{Right:} \xiion~in terms of sSFR. EELG1002 is found to be consistent with the highest sSFR and \xiion~sources identified in CEERS \citep{Whitler2024} and MIDIS \citep{Rinaldi2023} while also being higher than $2 < z< 5$ galaxies in VANDELS \citep{Castellano2023} and EIGER. This highlights how EELG1002 has ionization properties consistent with some of the most `extreme', bursty, and ionizing sources currently identified at $z > 5$ and provides a unique case study of low-$z$ analogs of high-$z$ galaxies.}
			\label{fig:xi_ion}
		\end{figure*}
		
		\subsection{Extreme \xiion~for Low-$z$ ELG but Reminiscent of Typical High-$z$ Galaxies}

		In previous sections, we find EELG1002 is characterized as having a highly energetic ISM with ionization parameters of $\log_{10} U = -2.23\pm0.06$ (\cigale) and $-1.95^{+0.06}_{-0.06}$ (\bagpipes) and a hard ionizing radiation field favoring both O$^{++}$ and Ne$^{++}$. Measurements of $n_e$ and $T_e$ in conjunction with the ionization parameter, as defined in Equation \ref{eqn:logU}, also show favorable conditions for increased ionizing photon production which, based on our results in \S\ref{sec:AGN}, is attributed to star-formation processes. We report measurements of the ionizing photon production efficiency and $1500$ \AA~magnitudes in Table \ref{table:SED} where we find EELG1002 has $\log_{10}\textrm{\xiionHtwo} = 25.74$ and $M_{UV} = -19.34$ using both \cigale- and \bagpipes-measured continuum. Despite the different IMF and stellar population models assumed in \cigale~and \bagpipes, both yield \xiion~are in $1\sigma$ agreement consistent with expectations for young, metal-poor stellar populations. However, we note that \xiion~can also be influenced by other factors such as metallicity, IMF, and stellar evolution effects.
	
		Both the \cigale~and \bagpipes-based \xiion~measurements are well above the canonical value expected of high-$z$ star-forming galaxy populations (e.g., \xiion~$\sim 10^{25.1 - 25.3}$ erg$^{-1}$ Hz; \citealt{Robertson2013}). EELG1002 also has elevated \xiion~relative to galaxies in the local Universe and at $z \sim 1$. Local LyC leakers are reported to range between \xiion$ \sim 10^{25.1 - 25.5}$ erg$^{-1}$ Hz \citep{Schaerer2016}. Compact star-forming galaxies at $0 < z < 1$ are found to range from $10^{24.5 - 25.5}$ erg$^{-1}$ Hz with a small fraction extending up to $10^{25.8}$ erg$^{-1}$ Hz consistent with EELG1002 \citep{Izotov2017}. We also find EELG1002 has higher \xiion~compared to typical $z \sim 2$ galaxies which report typical \xiion $\sim 10^{24.8 - 25.3}$ erg$^{-1}$ Hz \citep{Matthee2017,Shivaei2018}.
		
		Figure \ref{fig:xi_ion} shows \xiion~of EELG1002 versus \oiii+\hbeta~EW (\textit{left} panel) where we find it is consistent with the highest known EW emitters identified in CEERS at $5  < z < 8$  \citep{Chen2024} and $7 < z < 9$ \citep{Tang2023}, as well as past $z \sim 7$ \textit{Spitzer}/IRAC excess studies \citep{Endsley2021}. The \textit{right} panel of Figure \ref{fig:xi_ion} shows \xiion~versus sSFR where we use \hbeta~SFR along with stellar mass from \cigale~and \bagpipes~to measure sSFR, as reported in Table \ref{table:SED}. We find EELG1002 in both the \cigale~and \bagpipes-based measurement has \xiion~and sSFR also consistent with $7 < z < 9$ galaxies identified in CEERS \citep{Whitler2024} and MIDIS \citep{Rinaldi2023} and in the \textit{JWST} ERO results \citep{Sun2023}. Both the sSFR and \xiion~of EELG1002 are highly elevated relative to even $z \sim 2$ galaxy populations studied in VANDELS \citep{Castellano2023}. This suggests that EELG1002 is not only highly efficient in producing ionizing photons, but also is efficient at levels highly consistent with even some of the most starbursty, high \oiii+\hbeta~EW emitters currently being identified with \textit{JWST} in the $ z> 7$ Universe. In this regard, EELG1002 is an interesting case study of the ionizing properties of high-$z$ star-forming galaxies but within the low-$z$ Universe.

		\subsection{Recent Rapid Burst of Star Formation}
		\label{sec:sfh}
		
		\newcommand{\MorphologyTable}{\input{Morphology_table.table}}
		\begin{table}
			\centering
			\caption{Basic morphology properties from \texttt{pysersic} using \textit{HST}/ACS F814W imaging. The effective radius, $r_e$, and index, $n$, are based on the best-fit S\'ersic profile. Dynamical mass is measured using Equation \ref{eqn:dynamical_mass} with uncertainties also factoring in varying scaling factors between $1$ -- $5$ (\S\ref{sec:dynamical_mass}).}
			\label{table:morphology}
			\begin{tabular*}{\columnwidth}{@{\extracolsep{\fill}}lc}
				\hline
				Property & Measurement \\
				\hline
				\input{Morphology_table.table}
				\hline				
			\end{tabular*}
		\end{table}

        		\newcommand{\SEDFitProps}{\input{SED_fit_properties.table}}
		\begin{table*}
			\centering
			\caption{Best-Fit \cigale~and \bagpipes~stellar mass, SFR, and sSFR measurements. Star Formation Rate Surface Density, $\Sigma_\textrm{SFR}$, is measured using the effective radius from \texttt{pysersic} as shown in Table \ref{table:morphology} and defined by Equation \ref{eqn:sigma_SFR}. SFR from \bagpipes~are measured by taking the inferred SFH and integrating on the timescales listed below. Note that the \bagpipes~SFR(1 Myr) is constant up to 3 Myr. The SFR measurements for GMOS are based on H$\beta$ emission and cover a timescale of $\sim 3$ -- 10 Myr. The two measurements for GMOS sSFR are based on the \hbeta~SFR divided by the \cigale~stellar mass and \bagpipes~stellar mass in that order. \xiion~is measured using Equation \ref{eqn:xiion} and $M_{UV}$ is measured based on the best-fit SED continuum flux within a $1500\pm50$\AA~top-hat filter.}
			\label{table:SED}
			\begin{tabular*}{\textwidth}{@{\extracolsep{\fill}}lccc}
					\hline
					Property & \verb|Cigale| & \verb|Bagpipes| & GMOS \\
					\hline
					\input{SED_fit_properties.table}
					\hline				
			\end{tabular*}
		\end{table*}

		The ISM and ionization properties of EELG1002 suggest a stellar population that is dominated by young, massive stars capable of producing large quantities of ionizing photons and a harder ionizing radiation spectrum. This also indicates that EELG1002 is undergoing a recent, rapid increase/burst of star formation activity. As described in \S\ref{sec:SED}, we explore both parametric (delayed-$\tau$ + recent burst; \cigale) and non-parametric (\bagpipes) star formation history modeling in our SED fitting.

		Table \ref{table:SED} shows SFRs measured on different timescales where we find that both \cigale~and \bagpipes~measure about an order-of-magnitude increase in the SFRs from 100 to 10 Myrs. Despite the different modeling used, we find that both parametric and non-parametric SFH models are consistent with 100 Myr SFRs of $\sim 0.2$ -- $0.4$ \msol~yr$^{-1}$ and a rapidly increasing SFH reaching 10 Myr SFRs of $\sim 1.7$ -- $2.5$ \msol~yr$^{-1}$. Only \cigale~measures a 1 Myr SFR of $22.33\pm4.35$ \msol~yr$^{-1}$ as this is the default minimum time resolution in its SFH modeling, while \bagpipes~non-parametric SFH has a time resolution of 3 Myr (see \S\ref{sec:bagpipes}) with SFR if $5.39^{+0.22}_{-0.16}$ \msol~yr$^{-1}$. This is relatively consistent with the \hbeta~SFR of $7.7\pm0.4$ \msol~yr$^{-1}$ measured using \cite{Kennicutt1998} calibration assuming \cite{Chabrier2003} IMF which traces timescales of $\sim 1 - 10$ Myr (depending on the stellar population model). Overall, the SFRs suggest an intense, rapidly increasing burst of star-formation activity in recent times.
		
		For a low-mass galaxy such as EELG1002, a high SFR would be indicative of high specific SFR (sSFR)  placing it well above the SFR -- stellar mass correlation (`starburst'; e.g., \citealt{Speagle2014}). Table \ref{table:SED} includes the sSFR measured on 1, 10, and 100 Myr timescales along with \hbeta-measured sSFR assuming both the stellar mass measured from \cigale~and \bagpipes. As we saw for the SFRs, the sSFR also increases an order-of-magnitude from 100 to 10 Myr as well as from 10 to 1 Myr. We also find \hbeta-measured sSFR of $67.5\pm26$ Gyr$^{-1}$ (\cigale) and $28.1^{+16}_{-18}$ Gyr$^{-1}$ (\bagpipes) which would mean mass doubling times (inverse of the sSFR) $\sim  15$ -- $35$ Myr. This suggests that not only is EELG1002 undergoing a rapid burst of star formation activity, but is also rapidly building up its stellar mass content. 
		
		Star Formation Rate Surface Densities, $\Sigma_\textrm{SFR}$ are measured as:
		\begin{equation}
			\centering
			\Sigma_\textrm{SFR} = \frac{\textrm{SFR}}{2 \pi r_e^2}
			\label{eqn:sigma_SFR}
		\end{equation}
		and are shown in Table \ref{table:SED}. Given the compact nature of EELG1002 along with its high SFR, we find that $\Sigma_\textrm{SFR}$ on short timescales is quite high reaching $\sim 12.7$ M$_\odot$ yr$^{-1}$ kpc$^{-2}$ in the instantaneous, 1 Myr SFR measured with \cigale. The \hbeta-measured $\Sigma_\textrm{SFR}$ is somewhat lower at $4.4\pm0.3$ M$_\odot$ yr$^{-1}$ kpc$^{-2}$ and in better agreement with $3.1\pm0.2$ M$_\odot$ yr$^{-1}$ kpc$^{-2}$ measured with \bagpipes. However, all our measurements of $\Sigma_{SFR}$ are similar to galaxies with similar O32, stellar mass, and $U$ at $z > 3$ \citep{Reddy2023}. This suggests that star formation activity is quite compact in EELG1002 and may be another reason why properties such as \xiion, $T_e$, and $U$ are elevated given the concentration of recently formed hot, massive stars collectively releasing ionizing photons into the ISM.

		\section{Discussion}
            \label{sec:discussion}
		
		\subsection{Realistic Star Formation Histories or Outshining Effect?}
		The rapidly rising and intense star-formation history of EELG1002 without any past star-formation activity at higher lookback times does raise the question: \textit{does the star formation histories conform to our current framework of galaxy formation \& evolution or is this a result of an outshining effect?} The latter would suggest that the recently formed young stellar population is bright enough to `outshine' the old, mature stellar population formed at older lookback times. This would result in a biased SFH and an underestimation in stellar masses \citep{Narayanan2024}. In the case of EELG1002, we have near-IR constraints from CFHT/WIRCam and \textit{HST} WFC3/F140W to potentially constrain the older stellar population. Future IR observations (e.g., $> 1$ \micron~rest-frame) could provide better constraints on the old stellar population, if present.
			
		However, \textit{does such SFHs even fit within the framework of galaxy formation \& evolution?} To answer this question, we look at the Illustris simulation \citep{Genel2014,Vogelsberger2014,Vogelsberger2014_galpop,Nelson2015} for analogs of EELG1002. We search for analogs within TNG50, 100, and 300 in snapshot \#55 ($z = 0.82$) that have properties broadly consistent with EELG1002: SFR $> 3$ M$_\odot$ yr$^{-1}$, stellar mass between $10^7$ and $10^{8.7}$, and $ugriz$ magnitudes near $-19.5\pm1$ mag. We also limit the analogs to those that have not undergone a recent merger with another subhalo and also have gas-phase metallicities roughly consistent with EELG1002. 
		
		In total, we identify 2 strong candidates within TNG300-2 in snapshot \#55 with IDs of 182200 and 119294 and show the star-formation histories in Figure \ref{fig:analogs}. In the case of 182200, the subhalo was recently formed with a rapid increase in its star formation rates. This does not necessarily mean that such a galaxy has formation time at $z \sim 0.8$ rather the subhalo and all associated particles were resolved by this snapshot in the simulation. The SFR measured at the first snapshot (\#53; $z \sim 0.89$) was $0.04$ \msol~yr$^{-1}$ and quickly rose by almost two orders of magnitude to $3.1$ \msol~yr$^{-1}$ by $z \sim 0.82$ all within a $\sim 330$ Myr time frame. This is very much reminiscent of the rapid and intense burst of star-formation we find based on both parametric and non-parametric SFH modeling of EELG1002.
		
		On the other hand, 119294 is somewhat older with a resolved formation time extending back to $z \sim 1.1$ where it starts with a small initial burst of $\sim 0.26$ \msol~yr$^{-1}$ that quickly died out within $\sim 250$ Myr and remained inactive with no star-formation activity until $z \sim 0.89$. At this point, 119294 had a star formation rate of $0.06$ \msol~yr$^{-1}$ and rapidly increased in SFR up to 3.9 \msol yr$^{-1}$ by $z \sim 0.82$ very much similar to 182200. Overall, this suggests that the SFH we find for EELG1002 does conform with our current framework for galaxy formation \& evolution. Our comparison to Illustris analogs also suggests that EELG1002 may have undergone previous minor bursts of star-formation as well which could form a low-mass older population. Deep near-IR imaging and spectroscopy could shed light if such a population exists; however, based on the available evidence, EELG1002 has an SFH described as a rapid burst within a $\sim 10$ Myr timescale that is found to be `realistic' in this regard that EELG1002-like sources do show within large-scale cosmological simulations such as Illustris.

		\subsection{What is driving the intense star formation activity?}
		
		Given that such intense star formation activity is supported by simulations, a follow-up question that arises is: \textit{what is driving such intense, rapid star formation activity?} In order to have such star formation requires that a galaxy has a substantial amount of cold gas available and a pathway for the accretion of more cold gas to continuously refuel the reservoir. In the case of EELG1002, we find mass doubling times of $\sim 15$ -- $35$ Myr that would suggest a sizable amount of cold gas is available to cause such rapid stellar mass growth. The low gas-phase metallicity of EELG1002 would also suggest that cold and relatively untouched gas reservoirs are available and that, as we saw in the Illustris analogs, EELG1002 may be undergoing a first starburst phase. Otherwise, given its redshift, we would expect a higher gas-phase metallicity reflecting chemical enrichment from past star formation episodes. 
		
		The \textit{middle} panel of Figure \ref{fig:analogs} also supports the idea that gas fractions within sources like EELG1002 are quite elevated. In the two Illustris analogs, we find gas masses of $\sim 10^9$ \msol~which is $10$ -- 100 $\times$ higher than its stellar mass content corresponding to $>90$\% gas fractions. The low gas metallicities in the analogs shown in the \textit{bottom} panel of Figure \ref{fig:analogs} demonstrate the lack of chemical evolution given very little past star formation activity at earlier times. 
		
		Observationally, we find evidence for potentially high gas fractions within EELG1002. The dynamical mass of EELG1002 is $\sim (4.2^{+27.9}_{-4.2}) \times 10^9$ \msol~and represents the combination of stellar, gas, and dark matter within the galaxy. Although dynamical mass also includes the dust mass, we find based on Balmer Decrement that $E(B - V) \sim 0$ mag such that dust mass is most likely negligible. The dynamical mass relative to the stellar mass based on \cigale~and \bagpipes~suggests that $> 90$\% of the dynamical mass consists of both gas and dark matter consistent with the Illustris analogs suggesting a high gas fraction. Inversing the Kennicutt-Schmidt law \citep{Kennicutt1998_KS} and using the $\Sigma_{SFR}$ measurement based on \hbeta~emission (see Table \ref{table:SED}), we find an inferred gas mass of $\sim 10^9$ \msol~comparable to the dynamical mass. Sources at similar \oiii~luminosity as EELG1002 are also found to reside in halos with typical masses ranging between $\sim 10^{12.5}$ and $10^{13}$ \msol~\citep{Khostovan2018} which would suggest for deep gravitational potentials that would facilitate the inflow of gas. The Illustris analogs also reside in group halos with dark matter mass of $\sim 10^{12.2}$ and $10^{12.6}$ \msol~for 182200 and 119294, respectively. Although we do not observe any gas inflow features based on the GMOS spectra, enhanced pristine cold gas accretion mixing with the ISM could also reduce the gas-phase metallicity. 
		
		We conclude that the SFH of EELG1002 is not due to an outshining effect and that it conforms to our current framework of galaxy formation and evolution as seen by the Illustris analogs. EELG1002 is most likely undergoing a first bursty phase of star formation activity as expected in high-$z$ galaxies (e.g., \citealt{Cohn2018}) which can also explain the high \oiii~EW given the lack of past stellar mass growth (low continuum flux density). What is most likely driving the intense star formation activity is the availability of copious amounts of gas (high gas fractions) coupled with potentially residing in dark matter halos with masses sufficient to facilitate the inflow of cold gas to replenish the gas reservoirs. Inversing the Kennicutt-Schmidt law \citep{Kennicutt1998_KS} and measuring the gas-consumption timescale such sustained star formation can persist for $\sim 250$ Myr comparable to what is found for the Illustris analogs. Follow-up observations with ALMA could shed light on the amount of cold, molecular gas available within EELG1002, as well as how efficiently the gas is being converted into stars (e.g., star formation efficiency).

	\subsection{What becomes of sources like EELG1002?}

        \begin{figure}
		\centering
		\includegraphics[width=\columnwidth]{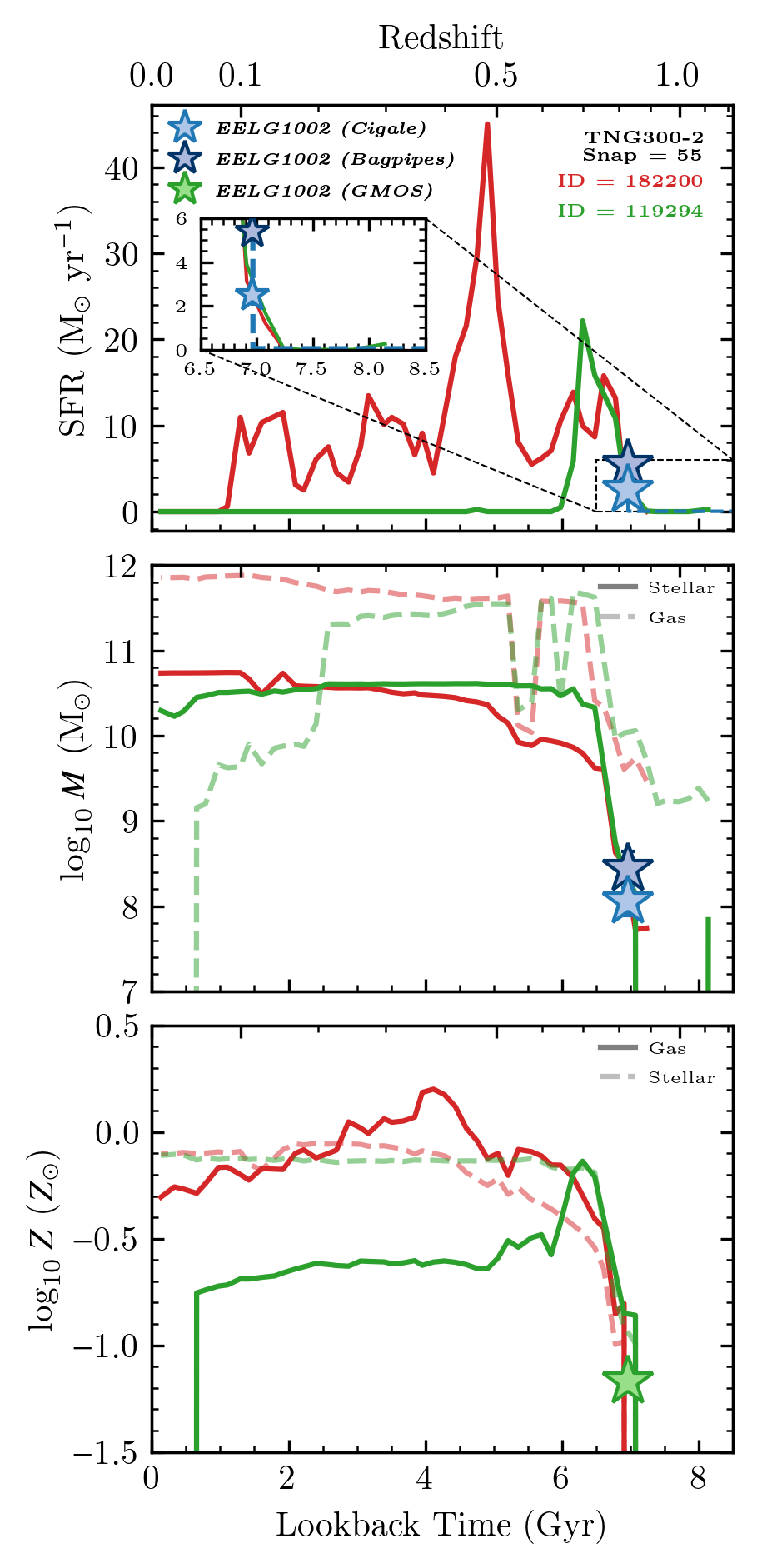}
		\caption{Illustris Analogs of EELG1002 with their respective star formation (\textit{top}), stellar and gas mass (\textit{middle}), and chemical enrichment (\textit{bottom}) histories. Both analogs show a rapid increase in star formation activity occurring at $z \sim 0.8$ signifying a first starburst phase where both are chemically unevolved and supported by high gas masses. Rapid chemical enrichment and stellar mass buildup occur within 100 -- 250 Myrs. Subsequent histories past $z < 0.7$ are different between both analogs owing to various mergers; however, both eventually become $10^{10 - 11}$ \msol~galaxies with solar-level metallicities. Therefore, EELG1002-like sources may potentially merge with more massive systems to become large structures in the local Universe. However, if EELG1002 remains isolated, then it could be the progenitor of a massive, compact quiescent galaxy (e.g., \citealt{Barro2013,Zolotov2015}).}
			\label{fig:analogs}
		\end{figure}
		
		Studies suggest that compact star-forming galaxies are the progenitors of compact, massive quiescent galaxies \citep{Barro2013,Zolotov2015}. We use the analogs identified in Illustris to map out the evolutionary path of EELG1002-like sources. Figure \ref{fig:analogs} shows the SFH (\textit{top}), gas and stellar mass growth (\textit{middle}), and gas and stellar chemical enrichment histories (\textit{bottom}). In both analogs, we find that the recent rise in star formation activity is part of an increasingly intense period of star formation. We will discuss the evolutionary path of each analog separately below with 119294 and 182200 shown in \textit{green} and \textit{red} in Figure \ref{fig:analogs}.

		\textit{119294}: This analog starts with a small, minor burst of star formation (\textit{top} panel inset of Figure \ref{fig:analogs}) at $z \sim 1.1$ (particle resolution limit in Illustris; not necessarily $t_{form}$) that dies out within $\sim 300$ Myr. This is followed by a single, large burst in star-formation activity starting at $z \sim 0.8$ that was sustained for 1 Gyr before eventually coming to a halt by $z \sim 0.5$ and lacking any significant star-formation activity down to $z \sim 0$. The system contained a high gas-to-stellar mass ratio at $z > 0.8$ that fueled the rapidly increasing SFH. By the next snapshot ($\sim 100$ Myr), 119294 nearly tripled its stellar mass from $2$ to $5.5 \times 10^8$ \msol~and maintained a gas mass of $\sim 10^{10}$ \msol. By $z = 0.73$ ($\sim 500$ Myr later), 1192924 experienced a merging event with a massive system which explains the rapid rise in SFH that persisted up to $z\sim 0.7$ followed by a decrease and eventual halt in star formation activity. The chemical enrichment of 119294 during the pre-merge star formation activity also shows gas metallicities doubling from $0.128$ to $0.205$ \zsol~and stellar metallicities from $0.111$ to $0.152$ \zsol~highlighting a rapid chemical enrichment period within $100$ Myrs. After the merger event, we find 119294 eventually becomes an essentially `dead'/quiescent massive galaxy ($\sim 10^{10}$ \msol) with very little gas mass and chemical enrichment consistent with $\sim \textrm{\zsol}$.
		
		\textit{182200}: The \textit{top} panel of Figure \ref{fig:analogs} shows a rapid increase in star formation activity starting at $z \sim 0.85$ (7.2 Gyr ago) where by $z \sim 0.82$ the SFR reaches $\sim 3$ \msol~yr$^{-1}$ with stellar mass of $3.2\times10^8$ \msol~and gas mass of $4 \times 10^9$ \msol, highlighting the gas-enriched environment that fuels the star formation activity. By the next snapshot ($100$ Myr later), the SFR increases to $13.2$ \msol~yr$^{-1}$ with a stellar mass of $4.2\times10^8$ \msol~and gas mass of $8.8\times10^9$ \msol, where the increase in gas mass indicates accretion to replenish the gas reservoir. 182200 experiences its first merging event by $z \sim 0.75$ followed by subsequent merging events corresponding to the peaks in its SFH (\textit{top} panel of Figure \ref{fig:analogs}) and by $z \sim 0.1$ eventually halts in star formation activity. The gas mass remains quite high by $z \sim 0$ and could indicate hot gas remains within this system. The chemical enrichment history pre-first merger shows a rapid increase within $\sim 250$ Myr from $0.04$ \zsol ($z \sim 0.85$) to $0.14$ \zsol ($z \sim 0.8$). After the merging events, we find 182200 shows an increase to $\sim$ \zsol~within 1 -- 1.5 Gyr after the initial burst. We find a similar chemical enrichment history when looking at stellar metallicities. By $z \sim 0$, 182200 has evolved into a chemically mature quiescent galaxy with stellar masses of a few times $10^{10}$ \msol. 
		
		Based on these two analogs alone, we find EELG1002 could most likely experience a merging event somewhere along its 7 Gyr history to the present-day and coalesce to become the massive, chemically-enriched galaxies that we see today. However, in the case that EELG1002 remains isolated (e.g., `field galaxy'), then it could potentially evolve in isolation into a massive, compact quiescent galaxy (e.g., \citealt{Barro2013,Zolotov2015}).

		\subsection{Lack of He\textsc{ii}4686\AA: Upper Limit in Ionizing Spectrum or Too Weak?}
		
		As we show in \S\ref{sec:ionization}, the ISM of EELG1002 is quite energetic and we find evidence for a hard ionizing radiation field given an elevated Ne3O2 ratio compared to O32. This suggests conditions for which Ne$^{++}$ is easily produced (requires 41 eV) compared to O$^{++}$ (requires 35 eV).  Coupled with the high \hbeta~EWs (signature of young stellar populations; e.g., \citealt{Fernandes2003,Levesque2013}), we would expect very high ionization potential lines such as He$^{++}$ which requires 54 eV. Past studies also suggest that galaxies having undergone a recent star-formation event and are dominated by low-metallicity young populations with high ionizing conditions are capable of producing \heii~emission (e.g., \citealt{Schaerer2003,Saxena2020,Berg2021}). Observations also suggest a potential increase in \heii4686\AA~emission with decreasing gas-phase metallicities at 12 + $\log_{10}$(O/H) $ < 8$ \citep{Senchyna2019}. 
		
		Despite the conditions favoring \heii~emission within EELG1002, we find no evidence of \heii4686\AA~emission in the GMOS spectra (\textit{top} panel of Figure \ref{fig:spectra_and_sed}). Although a lack of such emission could be an indication that there is an upper limit in the ionizing spectrum associated with EELG1002, we explore the possibility \heii4686\AA~emission was too faint to be observed with the GMOS observations (exposure time of 3600 seconds). \heii4686\AA~is included in the \texttt{Cloudy} models that were used within \bagpipes~where we find that \heii~is predicted to have a line flux of $\sim 2\times10^{-19}$ \cgsline~which does place it well below the detection limit ($\sim9.4\times10^{-18}$ \cgsline~about $4686\pm20$\AA~at $3\sigma$). However, studies have shown that photoionization models underestimate the contribution of \heii~emission (e.g., \citealt{Nanayakkara2019,Berg2021}). \cite{Senchyna2019} investigate the \heii4686\AA/\hbeta~line ratio as a function of $12+\log_{10}$(O/H) for a sample of six $z < 0.01$ extremely metal-poor ($Z < 0.1$ \zsol) galaxies and reported a gradient where \heii~emission was found to increase with decreasing gas-phase metallicity. Using their empirical measurement, we expect \heii4686\AA/\hbeta~in the range of $\sim 0.01$ to $0.04$ at $12+\log_{10}$(O/H)$\sim 7.5$ which would correspond to an \heii4686\AA~line flux of $\sim 1.87$ -- $7.48 \times 10^{-18}$ \cgsline~which is still well below our detection limit at the $3\sigma$ level. This would suggest that EELG1002 may have an ionizing spectrum that does extend to ionizing energies capable of producing He$^{++}$; however, our current data is not sensitive to such line fluxes to observe He{\sc ii} 4686\AA. 

		Overall, the lack of He{\sc ii}4686\AA~does not necessarily suggest an upper limit in the ionizing spectrum but more likely that the GMOS data was not sensitive to observe this specific \heii~line. However, this does raise the prospect of observing \heii1640\AA~which is typically 6 -- 8 $\times$ brighter than the 4686\AA~line (based on \pyneb~\texttt{getEmissivity} for He). Indeed if He{\sc ii}1640\AA~is observed within EELG1002, it would suggest very high ionizing conditions that are reminiscent of galaxies in the early Universe. Especially given that we find no evidence for X-ray binaries given the lack of X-ray detections (see \S\ref{sec:AGN}) and presence of Wolf-Rayet stars (lack of a blue WR bump; e.g., \citealt{Guseva2000}), then the main source could potentially be from a young, low-metallicity stellar population \citep{Saxena2020}. Observations with \textit{HST}/COS are needed to observe He{\sc ii}1640\AA~to confirm this.

		\subsection{Conditions for LyC Escape}
		\label{sec:fesc}
		
		EELG1002 potentially has conditions that could facilitate LyC escape. The compact size and $\Sigma_\textrm{SFR}$ in conjunction with the elevated $U$ and \xiion~suggest large quantities of ionizing photons are available and concentrated. This is also backed by the recent SFR, sSFR, and \hbeta~EW$\sim 404 - 473$\AA~suggesting the presence of a largely young stellar population is present within EELG1002. Mechanical feedback mechanisms such as stellar-driven or SNe-driven winds could result in the creation of low H{\sc i} column density channels allowing for Ly$\alpha$ and LyC escape (e.g., \citealt{Yang2016,Pucha2022,Reddy2022}). Such outflows are not seen within the GMOS spectra (e.g., asymmetric line profiles) but may be observable within rest-frame UV spectroscopy via P Cygni profiles around high ionization lines. Available \textit{GALEX}/FUV photometry also suggests a non-zero LyC escape fraction ($f_{esc}$) with measured \texttt{mag}$_\textrm{\texttt{AUTO}} = 26.25\pm0.34$ ($\sim 3\sigma$ detection). However, the \textit{GALEX}/FUV photometry could be contaminated by light redwards of the Lyman limit and suffer from blending issues due to the poor spatial resolution of \textit{GALEX} and the close angular proximity of two sources (see Figure \ref{fig:slitpos}). Without spatially-resolved UV imaging and/or deep UV spectroscopy, we can not directly measure $f_{esc}$ for EELG1002 although the conditions would suggest that LyC escape is present.
		
		Past studies have developed empirical calibrations to estimate $f_{esc}$ given observables. We use these indicators to infer $f_{esc}$ but note that various caveats are associated with each calibration for which we refer the reader to \cite{Choustikov2024} for a detailed overview. We first use the empirical O32 calibration of \cite{Faisst2016_O32} and infer $f_{esc} > 0.115$ at the 90\% confidence level. We next use the UV spectral slope, $\beta$, calibration of \cite{Chisholm2022} which was based on sources observed in the Low-$z$ Lyman Continuum Survey (LzLCS). In both \cigale~and \bagpipes~SED fitting, we find EELG1002 has $\beta \sim -2.5$ and corresponds to $f_{esc} \sim 0.08 - 0.21$ based on this calibration. Next, we use the $\Sigma_{SFR}$ calibration of \cite{Naidu2020} motivated by more compact, star-forming systems with high $\Sigma_\textrm{SFR}$ having higher $f_{esc}$ given feedback mechanisms forming low-density channels allowing for LyC escape. We infer $f_{esc} \sim 0.15 - 0.22$ using this calibration. Lastly, \cite{Choustikov2024} developed a combined $f_{esc}$ indicator that is dependent on several key properties related to LyC escape: $\beta$, $E(B-V)$, \hbeta~luminosity, $M_{UV}$, \textit{R23}, and O32. We find an inferred $f_{esc} \sim 0.13$ using this calibration. Overall, EELG1002 potentially has $f_{esc} \sim 0.1 - 0.2$ based on the above mentioned calibrations and future rest-frame UV observations are needed for confirmation.

		\subsection{Ideal Case of Reionization-Era Galaxies?}
		
		EELG1002 is a uniquely extreme source at $z \sim 0.8$ but may represent the conditions that existed in Reionization-era galaxies. Throughout this paper, we have noted how EELG1002 is similar to galaxies currently being observed in the $z > 5$ Universe by \textit{JWST}. We found that EELG1002 has \oiii+\hbeta~EW at fixed stellar mass even higher than typical star-forming galaxies observed in EIGER \citep{Matthee2023} and JADES \citep{Boyett2024}. Gas-phase metallicity, ionization parameter, and excitation state ($R23$) are also found to be similar to $z > 5$ galaxies. \xiion~at fixed \oiii+\hbeta~is highly consistent with $z \sim 7 - 9$ galaxies from CEERS \citep{Tang2023}. EELG1002 also has sSFR and \xiion~consistent with some of the most extreme systems identified in CEERS \citep{Whitler2024} and MIDIS \citep{Rinaldi2023}. The optical size of EELG1002 is $\sim 530$ pc and is consistent with the typical optical sizes of $7 < z < 9$ galaxies at similar $4800$\AA~rest-frame magnitudes ($-18.8$ mag; $\sim 400$ pc with scatter up to $\sim 700$ pc; \citealt{Yang2022_Lilan}). Overall, EELG1002 matches with $z > 5$ galaxies in many ways which highlights a key point that this system represents similar characteristics and properties as galaxies within the Epoch of Reionization. Future follow-up studies of EELG1002 can shed more light on the ionization and star formation processes in the context of what may have also occurred at high-$z$. Although EELG1002 is only a single object, the work presented here motivates for similar searches of high EW objects within archival datasets which can form statistically larger samples of low-$z$ `extreme' galaxies similar to typical star-forming galaxies observed at $z > 5$ and used to study Reionization Era-like galaxies in great detail.

		\section{Conclusions}
            \label{sec:conclusions}
		
		In this paper, we have presented a detailed analysis of a $z = 0.8275$ extreme emission line galaxy, EELG1002, which was identified as part of ongoing work of the COSMOS spectroscopic archive within 7 year old Gemini GMOS-S spectroscopic data. We use all available spectroscopic and photometric data to investigate the nature of this source in great detail. Our main results are:
		
		\begin{enumerate}[nolistsep,label=(\roman*)]
			\item EELG1002 has rest-frame \oiii+\hbeta~EW $\sim 3100$ to $3700$\AA~with stellar mass of $\sim 10^8$ \msol~depending on the SED fitting model assumed for the continuum flux density. This is found to be $\sim 32 - 36 \times$ higher than the typical \oiii+\hbeta~EW for $z \sim 0.8$ at similar stellar mass outlining the extremity and rarity of such a source. The \oiii+\hbeta~EW is also higher than typical $z > 3$ ELGs  \citep{Khostovan2016,Boyett2024} and even $z \sim 5 - 7$ galaxies \citep{Matthee2023} and comparable to known local LyC leakers.
			
			\item Strong \hbeta~emission suggests recent star formation rates of $7.7$ \msol~yr$^{-1}$ with mass doubling timescales of $\sim 15 - 35$ Myr. Combined with its compact size ($r_{eff} \sim 530$ pc; proper), we find EELG1002 has $\Sigma_{SFR} \sim 4.4$ \msol~yr$^{-1}$ kpc$^{-2}$. Star formation history modeling using spectrophotometric SED fitting (\cigale~and \bagpipes) also confirms the bursty star-forming nature of EELG1002 with no past star-formation activity at older lookback times.
						
			\item We find no clear evidence of an AGN component in EELG1002 given the lack of broad emission line features, X-ray detection, and the necessary \oiii/\hbeta~and stellar mass to fall within the AGN classification using MEx diagnostic. \oiii/\hbeta~ratio coupled with low metallicity from direct $T_e$ would suggest low \nii/\halpha~making EELG1002 also fall within the BPT star-forming classification. There may be a potential obscured AGN component given \textit{Spitzer}/IRAC photometry; however, it is blended with 2 nearby sources. Furthermore, a dust obscured AGN may be unlikely given that Balmer Decrements suggest $E(B-V) = 0$ mag.
			
			\item We find EELG1002 is metal poor with $12+\log_{10} (\textrm{O/H}) = 7.52\pm0.07$ ($Z_{gas} = 0.068^{+0.013}_{-0.009}$ \zsol) based on direct $T_e$ measurements using the auroral \oiii4363\AA~line. At the measured stellar mass, we find EELG1002 has gas-phase metallicity consistent with $z > 5$ galaxies. Other known low-$z$ analogs of high-$z$ galaxies show higher stellar mass and metallicity compared to EELG1002 which would suggest some past star-formation activity and chemical enrichment. However, the low metallicity and stellar mass of EELG1002 suggests a lack of chemical evolution and past star-formation activity such that EELG1002 may represent a galaxy undergoing a potential first bursty phase of star formation.
			
			\item Spectrophotometric SED fitting using both \cigale~and \bagpipes~show elevated ionization parameters with $\log_{10} U \sim -2.23$ and $-1.96$, respectively. \oiii/\oii~ratios also suggest highly energetic ISM conditions and elevated \neiii/\oii~at fixed \oiii/\oii~show evidence of a harder ionizing radiation field (e.g., more EUV photons).  The lack of \heii4686\AA~emission is due to the observations not going deep enough to detect the line rather than an upper limit in the ionizing spectrum at 54 eV. Deep rest-frame UV spectroscopic follow-up could potentially yield \heii1640\AA~which is typically $\sim 6 - 8 \times$ brighter than its 4686\AA~counterpart.
			
			\item EELG1002 has $\log_{10} \textrm{\xiion} \sim 25.74$ that is consistent with some of the most extreme and bursty star-forming galaxies observed at $z > 7$ based on \oiii+\hbeta~EW and sSFR. The elevated efficiency in producing ionizing photons within EELG1002 may be attributed to a combination of elevated SFR, compact size, low metallicity, and highly energetic ISM condition with evidence of a harder ionizing radiation field. 
			
			\item Although we find a $3\sigma$ detection within \textit{GALEX}/FUV that would suggest LyC escape, we note that the \textit{GALEX} spatial resolution is poor and the $3\sigma$ detection may be contaminated by 2 nearby sources. Using multiple empirical calibrations, we find that EELG1002 may have $\sim 10 - 20\%$ LyC escape fraction.

			\item Both parameteric and non-parametric SFH modeling of EELG1002 suggest for a recent burst of star formation with no past activity at older lookback times. Analogs identified in Illustris-TNG suggest such SFHs fall within our current framework of galaxy formation \& evolution where intense star formation occurs followed by rapid chemical enrichment and stellar mass buildup. This is supported by high gas masses in the simulations. Observationally, we find EELG1002 has dynamical mass of $\sim 10^9$ \msol~which is an order-of-magnitude higher than its stellar mass suggesting high gas masses similar to what is found in the Illustris analogs.
		\end{enumerate}
		
		EELG1002 provides for an interesting and unique case of a low-$z$ galaxy with properties highly consistent with even some of the extreme star-forming galaxies currently being observed at $z > 5$ with \textit{JWST}. The \oiii+\hbeta~EW alone for EELG1002 is record-breaking for a star-forming galaxy at its redshift and stellar mass highlighting not only the extremity but rarity of this source. More importantly, we have demonstrated how such a source at low-$z$ can be used to uncover details on the ionizing and star-formation properties \& processes expected to occur in the high-$z$ Universe by using all available evidence and referring to large hydrodynamical simulations for support. This work also emphasizes the importance of archival datasets that are currently being processed and analyzed as part of on-going work in developing the COSMOS Spectroscopic Archive where unpublished data can present surprising scientific discoveries such as EELG1002. Future next-generation surveys planned with \textit{Euclid} and \textit{Roman} will find many EELG1002-like systems given the wide areal coverage resulting in large comoving volumes needed given the rarity (e.g., low number densities) of such sources.

		\section*{Acknowledgements}
		
		We thank the anonymous reviewer for their feedback and suggestions that enhanced the quality of this work. AAK thanks Yuichi Harikane, Erini Lambrides, Brian Lemaux, Keunho Kim, Jed McKinney, Mainak Singha, and Ryan Sanders for useful discussions. AAK also thanks all participants of the \href{https://noirlab.edu/science/events/websites/first-gigayears-2024}{NOIRLab First Gigayear(s) Conference} at Hilo, Hawaii and the \href{https://www.stsci.edu/contents/events/stsci/2024/april/recipes-to-regulate-star-formation-at-all-scales}{STScI Spring Symposium 2024: ``Recipes to Regulate Star Formation at All Scales: From the Nearby Universe to the First Galaxies''} for productive comments, feedback, and discussions. 

        This material is based upon work supported by the National Science Foundation under Grant No. 2009572.
		
		Based on observations obtained at the international Gemini Observatory, a program of NSF NOIRLab, which is managed by the Association of Universities for Research in Astronomy (AURA) under a cooperative agreement with the U.S. National Science Foundation on behalf of the Gemini Observatory partnership: the U.S. National Science Foundation (United States), National Research Council (Canada), Agencia Nacional de Investigaci\'{o}n y Desarrollo (Chile), Ministerio de Ciencia, Tecnolog\'{i}a e Innovaci\'{o}n (Argentina), Minist\'{e}rio da Ci\^{e}ncia, Tecnologia, Inova\c{c}\~{o}es e Comunica\c{c}\~{o}es (Brazil), and Korea Astronomy and Space Science Institute (Republic of Korea).
		
		\textit{Software:} \texttt{astropy} \citep{astropy:2013,astropy:2018,astropy:2022}, \texttt{numpy} \citep{harris2020array}, \pyqsofit~\citep{Guo2018,Shen2019}, \pyneb~\citep{Luridiana2015}, \texttt{pysersic} \citep{Pasha2023}, \pypeit~\citep{Prochaska2020, Prochaska_zenodo}, \cigale~\citep{Boquien2019,Yang2022}, \bagpipes~\citep{Carnall2018}

		\section*{Data Availability}
		
		All raw data is publicly available and can be found within the Gemini Science Archive by searching for GS-2017A-FT-9 under Program ID. The associated pypeit reduction files can be found at \href{https://github.com/akhostov/EELG1002}{GitHub repository} as well as within \dataset[doi:10.5281/zenodo.16990022]{https://doi.org/10.5281/zenodo.16990022} and are available for public use.
		
		
            \appendix
		\begin{table*}
			\caption{\texttt{Cigale} parameters used in the SED fitting process}
			\label{table:cigale}
			\centering
			\begin{tabular*}{\textwidth}{@{\extracolsep{\fill}}lcc}
				\hline
				Parameter & Symbol & Value \\
				\hline
				\textbf{Delayed Exponential Star Formation History with Burst} & & \\
				$e$--folding time of the main stellar population model (Gyr)     & $\tau_\textrm{main}$ & 0.05, 0.1, 0.25, 0.5, 0.75, 1.0, 2.5, 5.0, 7.5 \\
				Age of the main stellar population in the galaxy (Gyr)                 & $t_\textrm{main}$       & 0.05, 0.1, 0.25, 0.5, 0.75, 1.0, 2.5, 5.0, 6.6 \\
				$e$--folding time of the late starburst population model (Myr) & $\tau_\textrm{burst}$ & 1, 5, 10, 50, 100 \\
				Age of the late burst (Myr)                                                                & $t_\textrm{burst}$      & 1, 5, 10, 25, 50 \\
				Mass fraction of the late burst population                                    & $f_\textrm{burst}$      &  Uniform(0,1) steps of 0.1 \\
				\hline
				\textbf{Stellar Population Synthesis Model} \citep{Bruzual2003}         &                                        &  \\
				Initial mass function																		 &  IMF          					   & \citet{Chabrier2003} \\
				Stellar Metallicity ($Z_\odot$) 													  & $Z_\star$                      & 0.005, 0.02, 0.2, 0.4, 1. \\
				\hline
				\textbf{Nebular Emission Line Spectrum} (\texttt{Cloudy v13.01}; \citealt{Ferland1998,Ferland2013}) & & \\
				Ionisation parameter																	   & $\log_{10} U$	              & Uniform(-3,-1) steps of 0.1 \\		 
				Gas Metallicity ($Z_\odot$)															  & $Z_g$         & 0.05 (fixed) \\
				Electron Density (cm$^{-3}$)														& $n_e$							  & 1000 \\
				Fraction of Lyman Continuum photons escaping the galaxy      & $f_{esc}$					 & 0.0 \\
				Fraction of Lyman Continuum photons absorbed by dust          & $f_{esc,\rm{dust}}$    & 0.0 \\
				Emission Line Widths (km s$^{-1}$)											   &  $\Delta v_{lines}$       & 400 (fixed) \\
				\hline
				\textbf{Dust Attenuation Model} \citep{Calzetti2000}    &  &\\ 
				Reddening of the nebular lines light for young \& old population (mag) & $E(B-V)_l$ & 0.0 (fixed) \\
				Ratio of $E(B-V)_l$ to Stellar Continuum Reddening (mag) & $f_{E(B-V)}$ & 1.0 (fixed) \\
				Central wavelength of the UV bump (\AA)									  &     $\lambda_{UV,b}$   & 2175 (fixed) \\
				Width (FWHM) of the UV bump (\AA)											& $\Delta\lambda_{UV,b}$ & 350 \\
				Amplitude of the UV bump (3: Milky Way)									  &        $I_{UV,b}$ 				 & 0\\
				Slope delta of the power law modifying the attenuation curve &        -- 							  & 0.0 \\
				Extinction law to use for attenuating the emission lines flux    &        --                              & SMC \citep{Pei1992} \\
				Total-to-selective extinction ratio for extinction curve applied to emission lines & $A_V / E(B-V)$ & 2.93 \\
				\hline
				\textbf{Dust Emission Model} \citep{Draine2014} & & \\
				Mass fraction of PAH																	&       $q_\mathrm{PAH}$			&  2.50 \\
				Minimum Radiation Field																 &       $U_\mathrm{min}$            &   1.0 \\
				Power law slope $dU/dM \propto U^\alpha$								 &       $\alpha$               &    2.0 \\
				Fraction illuminated from $U_\mathrm{min}$ to $U_\mathrm{max}$ 				  &          $\gamma$           &    0.1 \\
				\hline
			\end{tabular*}
		\end{table*}

		\begin{table*}
			\caption{\bagpipes~parameters used in the SED fitting process. Note that \bagpipes~assumes }
			\label{table:bagpipes}
			\centering
			\begin{tabular*}{\textwidth}{@{\extracolsep{\fill}}lcc}
				\hline
				Parameter & Range & Prior \\
				\hline
				\textbf{Continuity Non-Parametric Star Formation History Model} \citep{Leja2019} & & \\
				Mass Formed in Each Time Bin ($\log_{10} M/$M$_\odot$)  & [1,13] & Uniform\\
				Stellar Metallicity ($Z_\odot$)               &	 $5\times10^{-4}$, 2]    & 	Uniform \\
				Time Bin Edges (Myr) & 0, 3, 10, 30, 100, 300, 1000, 3000, 6000 & -- \\
				\hline
				\textbf{Stellar Population Synthesis Model} (\texttt{BPASS v2.2.1} \citealt{Stanway2018})       &                                        &  \\
				Initial mass function -- Broken Power Law (Upper Mass: 300 \msol)			&  IMF          					   & -- \\
				\hline
				\textbf{Nebular Emission Line Spectrum} (\texttt{Cloudy v17.03}; recomputed for $n_e = 800$ cm$^{-3}$) & & \\
				Ionisation parameter ($\log_{10} U$)	              & [$-4$,$-1$] &  Uniform \\		 
				Gas Metallicity ($Z_\odot$)									&   0.065     & Fixed \\
				\hline
				\textbf{Dust Attenuation Model} \citep{Calzetti2000}    &  &\\ 
				$V$-band dust attenuation ($A_V$; mag) & 0.0  & Fixed \\
				$E(B-V)_\mathrm{nebular}/E(B-V)_\mathrm{stellar}$ & 1.0 & Fixed \\
				Slope delta of the power law modifying the attenuation curve &  0.0 & Fixed \\
				\hline
				\textbf{Dust Emission Model} \citep{Draine2014} & & \\
				Mass fraction of PAH ($q_\mathrm{PAH}$)			&  2.50 & Fixed \\
				Minimum Radiation Field ($U_\mathrm{min}$)        &   1.0 & Fixed \\
				Fraction illuminated from $U_\mathrm{min}$ to $U_\mathrm{max}$ ($\gamma$)           &    0.1  & Fixed\\
				\hline
			\end{tabular*}
		\end{table*}
		

    	\bibliographystyle{aasjournal}
		\bibliography{EELG_OIII_GMOS} 

	\end{document}